\documentclass[10pt,letterpaper]{article}
\usepackage[top=0.85in,left=2.75in,footskip=0.75in]{geometry}

\usepackage{amsmath,amssymb}
\usepackage[T1]{fontenc}
\usepackage[utf8]{inputenc}
\usepackage{changepage}

\usepackage{textcomp,marvosym}

\usepackage{cite}
\usepackage[normalem]{ulem}
\useunder{\uline}{\ul}{}
\usepackage{nameref,hyperref}

\usepackage[right]{lineno}

\usepackage[nopatch=eqnum]{microtype}
\DisableLigatures[f]{encoding = *, family = * }

\usepackage[table]{xcolor}

\usepackage{array}

\newcolumntype{+}{!{\vrule width 2pt}}

\newlength\savedwidth

\usepackage{algorithm}
\usepackage{algorithmic}

\usepackage{multirow}
\usepackage{multicol}

\raggedright
\usepackage[aboveskip=1pt,labelfont=bf,labelsep=period,justification=raggedright,singlelinecheck=off]{caption}

\makeatletter
\renewcommand{\@biblabel}[1]{\quad#1.}
\makeatother

\usepackage{lastpage,fancyhdr,graphicx}
\usepackage{epstopdf}
\fancyheadoffset[L]{2.25in}
\fancyfootoffset[L]{2.25in}
\begin{document}
\vspace*{0.2in}

% Title must be 250 characters or less.
\begin{flushleft}
{\Large
\textbf\newline{Bayesian uncertainty-aware deep learning with noisy labels: Tackling annotation ambiguity in EEG seizure detection} % Please use "sentence case" for title and headings (capitalize only the first word in a title (or heading), the first word in a subtitle (or subheading), and any proper nouns).
\\ 
}

\vspace{0.1cm}
{\small Short title: Bayesian uncertainty-aware deep learning for handling noisy seizure labels in EEG}
\newline

% Insert author names, affiliations and corresponding author email (do not include titles, positions, or degrees).
Deeksha M. Shama \textsuperscript{1,2*},
% Name2 Surname\textsuperscript{2\Yinyang},
% Name3 Surname\textsuperscript{2,3\textcurrency},
% Name4 Surname\textsuperscript{2},
% Name5 Surname\textsuperscript{2\ddag},
% Name6 Surname\textsuperscript{2\ddag},
Archana Venkataraman\textsuperscript{1,2}
% with the Lorem Ipsum Consortium\textsuperscript{\textpilcrow}
\\
\bigskip
\textbf{1}  Department of Electrical and Computer Engineering, Johns Hopkins University, Baltimore, MD, United States of America 
\\
\textbf{2} Department of Electrical and Computer Engineering, Boston University, Boston, MA, United States of America  
\\
\bigskip

% Insert additional author notes using the symbols described below. Insert symbol callouts after author names as necessary.
% 
% Remove or comment out the author notes below if they aren't used.
%
% Primary Equal Contribution Note
%\Yinyang These authors contributed equally to this work.

% Additional Equal Contribution Note
% Also use this double-dagger symbol for special authorship notes, such as senior authorship.
%\ddag These authors also contributed equally to this work.

% Current address notes
%\textcurrency Current Address: Dept/Program/Center, Institution Name, City, State, Country % change symbol to "\textcurrency a" if more than one current address note
% \textcurrency b Insert second current address 
% \textcurrency c Insert third current address

% Deceased author note
%\dag Deceased

% Group/Consortium Author Note
%\textpilcrow Membership list can be found in the Acknowledgments section.

% Use the asterisk to denote corresponding authorship and provide email address in note below.
* dshama1@jhu.edu

\end{flushleft}
% Please keep the abstract below 300 words
\section*{Abstract}
 Deep learning is advancing EEG processing for automated epileptic seizure detection and onset zone localization, yet its performance relies heavily on high-quality annotated training data. However, scalp EEG is susceptible to high noise levels, which in turn leads to imprecise annotations of the seizure timing and characteristics. This “label noise” presents a significant challenge in model training and generalization. In this paper, we introduce Bayesian UncertaiNty-aware Deep Learning (BUNDL), a novel algorithm that informs a deep learning model of label ambiguities, thereby enhancing the robustness of seizure detection systems. By integrating domain knowledge into an underlying Bayesian framework, we derive a novel KL-divergence-based loss function that capitalizes on uncertainty to better learn seizure characteristics from scalp EEG. Thus, BUNDL offers a straightforward and model-agnostic method for training deep neural networks with noisy training labels that does not add any parameters to existing architectures. Additionally, we explore the impact of improved detection system on the task of automated onset zone localization. We validate BUNDL using a comprehensive simulated EEG dataset and two publicly available datasets collected by Temple University Hospital (TUH) and Boston Children’s Hospital (CHB-MIT). Results show that BUNDL consistently identifies noisy labels and improves the robustness of three base models under various label noise conditions. { We also conduct ablation experiments on uncertainty quantification, evaluate cross-site generalizability to Siena EEG dataset, and quantify computational cost of all methods.} Furthermore, we demonstrate that BUNDL improves seizure onset zone localization accuracy. Ultimately, BUNDL presents as a reliable method that can be seamlessly integrated with existing deep models used in clinical practice, enabling the training of trustworthy models for epilepsy evaluation.

%\linenumbers

% Use "Eq" instead of "Equation" for equation citations.
\section*{Introduction}

Epilepsy, a neurological disorder marked by recurring seizures, often necessitates further evaluation particularly in drug-resistant cases to guide treatment, such as surgical resection. Scalp EEG is the primary modality for assessing epilepsy, involving prolonged monitoring to detect epileptiform discharges indicative of seizure timing(s) and location(s).  EEG also presents as a natural choice for presurgical planning complimented with MRI due to its non-invasiveness, affordability, and accessibility.  However, EEG signals are complex and artifact-prone, making manual review both labor-intensive and error-prone. To address these challenges, computer-aided models have been developed, with deep learning models  demonstrating remarkable success in seizure characterization. % detection and characeterization

Despite this progress, most deep learning models for seizure detection are based on a supervised learning paradigm, which uses a \textit{training dataset} of EEG recordings and ``ground truth" labels of seizure activity annotated by clinicians. Promisingly, deep learning models are highly capable of extracting relevant features under the assumption that the provided ``ground truth" labels are accurate. Unfortunately, clinician-annotated labels of seizure activity can suffer from manual error and poor inter-rater reliability. Thus, the provided annotations are only approximately correct, thus availing \textit{noisy labels} for training deep models.  Research shows that inter-rater agreement among clinicians interpreting EEG data hovers around 60\%, underscoring the inconsistency in annotations~\cite{halford2013standardized, halford2015inter}. Another study attributed this variability largely to differing decision thresholds among clinicians~\cite{jing2020interrater} where a lower threshold is adopted to minimize the risk of missing seizures, leading to over-segmentation of seizures and misdiagnosis  of epilepsy~\cite{amin2019role}. When AI models are trained without considering such label ambiguities, they may learn misleading features, resulting in poor generalizability. In fact, previous studies have suggested that the performance of AI models may plateau on widely used datasets %like the Temple University Hospital Corpus 
due to these ambiguities~\cite{gemein2020machine}. Thus, there is a need for robust, noise-aware deep learning frameworks that can enhance the reliability of automated seizure detection.% systems.

\subsection*{Learning with label noise}
%one column full

The challenge of training with ambiguous or inaccurate labels is gaining traction within the deep learning community. Early approaches focus on eliminating noisy samples from the training dataset using prior information about the task. For example, ~\cite{chen2019understanding,park2023robust} use a pretrained network to identify samples with highest loss as noisy whereas Confidence Learning method~\cite{northcutt2021confident} use confusion matrices to detect noisy samples. The model is then retrained only on clean samples. %Similarly, \cite{park2023robust} incorporates a neighborhood constraint to ensure smooth sample rejection by leveraging data correlations.  
Such ``hard pruning" strategies are entirely dependent on the accuracy of the pretrained network and can erroneously keep/reject samples. In contrast, ``soft pruning" reweights samples during training to control how much the model learns from each sample based on its noise level%, while also addressing the class imbalance common in medical datasets
~\cite{xue2019robust}. The drawbacks of pruning strategies are that the ``prior information"  is often unavailable and in complex data like EEG, all samples may appear  noisy, leading to the potential removal of key samples. %or, in an extreme case, the rejection of the entire dataset.

The second set of approaches introduce novel architectural changes to deep neural networks that aim to decipher a set of ``clean labels" from the noisy training data. For instance, the work of \cite{goldberger2016training} elegantly addresses the learning challenge by adding a noise adaptation layer on top of an existing architecture to model the transition from clean to noisy labels. %and draws inspiration from the expectation-maximization (EM) estimation procedure for inferring posterior probability distributions. 
Other works extend this by using confusion matrix-guided label transition estimation and adding optimization constraints to the loss function for improved robustness~\cite{sukhbaatar2014training, zhang2021learning}. However, these architecture changes add new parameteres at a scale of square of the number of predicted classes. %, $O(n^2)$. 
They also require a computationally intensive multi-stage training, complicating their practical application~\cite{wen2024histopathology,xu2022anti,chen2024label,hailat2018teacher,xiao2015learning}. 

Yet another strategy involves the development of novel loss functions to algorithmically learn the clean label posterior. For example, self-supervised methods~\cite{huang2020self} update the class confidence based on past predictions; however, they assume class-conditional noise and often overlook confounding factors from the input, which is common in complex data like EEG. Unsupervised methods, such as \cite{arazo2019unsupervised}, employ mix-up data augmentation and bootstrapping to estimate clean labels based on the principle that a linear combination of inputs should imply a linear combination in class probabilities. However, the assumption of linear data variations may not be appropriate for complex modalities like EEG. Assumption-free approaches offer a more flexible way to handle label noise by creating novel loss functions~\cite{li2023dynamics,zhang2018generalized}, but they  struggle to model input data-dependent label transitions, crucial in complex EEG data.

\subsection*{Automated seizure detection}

Automated seizure detection has been a focus of research for over three decades, traditionally following a three-stage pipeline:~(i) signal preprocessing and segmentation into short pseudo-stationary time windows, (ii)~extraction of discriminatory features, and (iii)~classifying each window as baseline or seizure activity with various machine learning methods~\cite{samiee2015epileptic,joshi2023spatiotemporal,liu2023epileptic}. The first stage typically applies filtering to remove noise while preserving relevant information. Machine learning-based denoising~\cite{qiu2018denoising} has also been used to boost prediction accuracy. However, these methods mainly address signal quality but \textbf{\textit{not}} ambiguities in the annotated seizure timings.

Feature extraction was originally hand-crafted, using methods such as spike counts~\cite{joshi2023spatiotemporal}, signal statistics (e.g., power, maximum amplitude)\cite{wu2019automatic}, and more complex non-linear, frequency, wavelet, and graph-theoretic analyses\cite{craley2018novel, liu2023epileptic, samiee2015epileptic, zandi2010automated, akbarian2020framework}. However, these features often failed to generalize beyond training datasets and struggled to capture patient-specific information, as inter-ictal artifacts could resemble seizures.
Recently, deep learning has emerged as a powerful alternative. Convolutional Neural Networks (CNNs)\cite{hassan2022epileptic, nogay2021detection, wagh2020eeg} and Graph Convolutional Networks (GCNs)\cite{chen2021multi} effectively model spatial patterns, while Recurrent Neural Networks (RNNs) and Long Short-Term Memory (LSTM) networks~\cite{guler2005recurrent, vidyaratne2016deep, hu2020scalp} capture temporal dynamics. More advanced models, including Temporal Graph Convolutional Networks (TGCNs)\cite{covert2019temporal}, Transformers\cite{m2023deepsoz, pedoeem2022tabs}, and hybrid CNN-LSTM architectures~\cite{craley2021automated}, have shown strong performance by learning complex spatio-temporal relationships.  Thus, deep networks have significantly improved automated seizure detection \cite{dash2024review}.

%These features are fed into variety of classification algorithms such as support vector machines~\cite{liu2023epileptic,samiee2015epileptic,nesaei2014real}, logistic regression~\cite{samiee2015epileptic}, Bayes~\cite{samiee2015epileptic}, random forests and decision trees~\cite{liu2023epileptic,joshi2023spatiotemporal}, and neural networks~\cite{samiee2015epileptic}. While they represent seminal contributions to the field, traditional approaches  often fail to  generalize to unseen patients. 

However, all of these advanced architectures are trained in a supervised manner in the final classification step assuming the correctness of provided ``ground truth" labels of the seizure interval.  This involves minimizing a cross-entropy loss (CEL) between the model predictions and the clinician-provided training labels. By design, this training strategy cannot handle ambiguities in the annotated seizure intervals. This assumption has contributed to poor generalization across heterogeneous patient cohorts~\cite{gemein2020machine}. {Prior work in automated seizure detection has made great strides in quantifying model uncertainty~\cite{borovac2022calibration,li2022eegunc}, improving model training using annotations from multiple clinicians~\cite{saab2020weak,borovac2022ensemble}, and test-time adaptation to reduce false alarms and enhance generalization~\cite{segal2023utilizing, mao2023online}. However, these approaches do not address the challenge of \textit{learning from noisy clinical labels} during model development without additional manual effort or computational overhead.}

{Uncertainty quantification distinguishes between epistemic uncertainty (limitations of the model itself) and aleatoric uncertainty (ambiguity inherent to the EEG signal)~\cite{de2023uncertainty}. Epistemic uncertainty has been estimated using Monte-Carlo dropout~\cite{gal2016dropout,jiahao2024uncertainty}, deep ensemble prediction variability~\cite{srivastava2014dropout,borovac2022ensemble}, and Bayesian neural networks. In contrast, aleatoric uncertainty, which captures inherent data ambiguity, is often modeled in a Bayesian framework by making explicit assumptions about the data via Gaussian or Laplacian priors~\cite{deng2023eeg}. While useful to compensate for noise and uncertainty in the EEG signals, these methods do not consider whether \textit{the supervisory labels} are accurate. As an alternative, we note that label noise directly influences model confidence, affects the learned softmax probability distributions, and may distort both epistemic and aleatoric estimates~\cite{algan2020label,goel2021robustness}. Thus, we hypothesize that the  uncertainty profiles can serve as a proxy for label noise. To our knowledge, this connection has not been previously examined.}

\subsection*{Our contributions}
%half column 

{We present BUNDL (Bayesian Uncertainty-aware Deep Learning), an automatic approach for handling data-dependent label noise in EEG-based seizure detection. Originally presented in \cite{shama2024uncertainty} as a method to compensate for over-segmented seizures by clinicians (a type of label noise), BUNDL is now expanded as a novel training strategy, with the goal of informing deep networks of various types of  label noise. In contrast to existing methods, BUNDL is specifically tailored to the challenges of EEG annotations with the following novel contributions in this manuscript:} 
    \begin{enumerate}
        \item {\textbf{Uncertainty-driven label correction using Bayesian graphical models:} We introduce a probabilistic training framework that models uncertainty-informed label transitions to handle multiple forms of label noise. Through a Bayesian formulation inspired from~\cite{shama2024uncertainty}, our framework called BUNDL derives a novel loss function that adjusts for noisy annotations while estimating seizure likelihoods using different uncertainty-quantification methods. The approach remains fully automatic, requires only clinician-provided labels (which might be noisy), and can flexibly incorporate multi-annotator information when available.}
        
        \item {\textbf{Model-agnostic algorithms with convergence analyses:} We provide  algorithms that can be integrated into any deep neural network architecture to mitigate the impact of noisy training labels. We evaluate our BUNDL framework across three noise settings: symmetric noise affecting both seizure and non-seizure labels, and two asymmetric noise settings that target each class separately. We show that BUNDL generalizes across architectures and noise scenarios, and it allows domain knowledge to be injected when appropriate.}
        
        \item {\textbf{Simulated EEG benchmarks for noisy-label evaluation:} Using SEREEGA, we propose a new standardized EEG simulation and noisy-label generation pipeline to enable controlled experimentation, reproducibility, and more rigorous development of noise-robust algorithms.}
        
        \item {\textbf{Comprehensive evaluation of performance and generalization:} We demonstrate the value of BUNDL using three deep network architectures applied to a simulated dataset  under controlled noise settings and three real-world datasets: a TUH EEG dataset \cite{shah2018temple} for model development and quantitative evaluation, the CHB-MIT dataset \cite{guttag2010chb} for real-world seizure detection evaluation, and the Siena dataset \cite{detti2020siena} for out-of-distribution testing. We also benchmark the generalization of TUH-trained models reported in \cite{shama2024uncertainty} on the unseen Siena dataset to assess robustness in cross-site generalization.}
        
        \item {\textbf{Expanded limitations and future work:} We provide an expanded discussion on robustness, computational considerations, and avenues for future work, including deeper integration of uncertainty measures to improve learning with noisy labels and enhance reliability.}
    \end{enumerate}

\section*{Materials and methods}
\subsection*{BUNDL: Bayesian Uncertainty-aware Deep Learning}

\begin{figure}[!h]
    %\centering
    \includegraphics[width=0.95\linewidth]{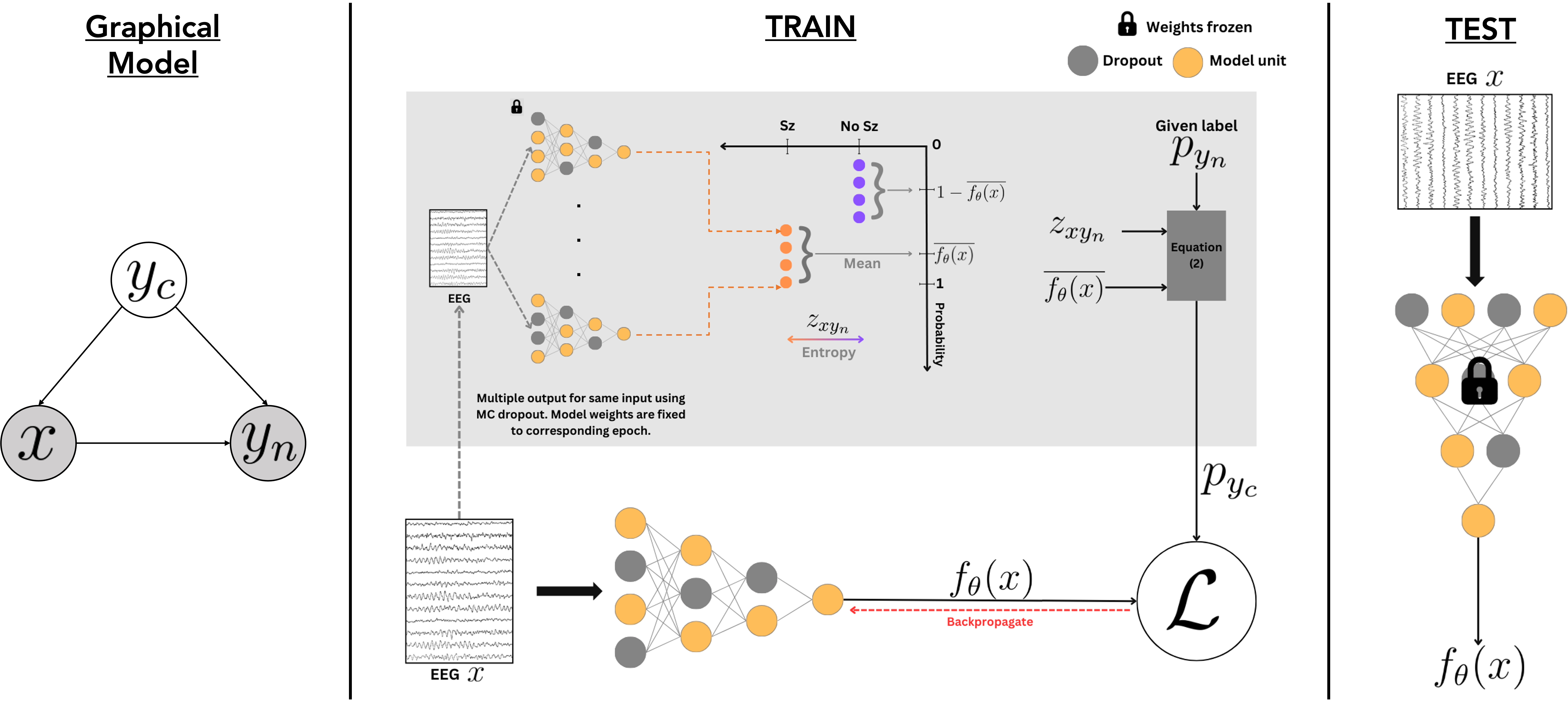}
    \caption{\textbf{Overall pipeline of BUNDL.}}
    \textbf{Left}: Graphical model depicting the relationships between random variables $x$ (EEG), $y_c$ (clean label), $y_n$ (noisy label). \textbf{Middle}: Overall training strategy of BUNDL. For every input EEG sample, we estimate uncertainty using MC dropout, as shown in the top pane. This information augments the standard prediction and backpropagation procedure via the BUNDL loss function, which is depicted in the bottom pane. \textbf{Right}: Testing pipeline for evaluating clean label prediction post training. 
    \label{train}
\end{figure}

% Place figure captions after the first paragraph in which they are cited.
% \begin{figure}[!h]
% \caption{{\bf Bold the figure title.}
% Figure caption text here, please use this space for the figure panel descriptions instead of using subfigure commands. A: Lorem ipsum dolor sit amet. B: Consectetur adipiscing elit.}
% \label{fig1}
% \end{figure}
Our Bayesian UNcertainty aware Deep Learning (BUNDL) framework  distinguishes between noisy and clean labels via the graphical model illustrated in Fig~\ref{train}(left). Specifically, we model the observed sample (in our case short EEG windows), represented by the random variable $x$, as dependent on the \textit{unobserved} clean labels $y_c$. Formally, $y_c$ is a binary random variable that indicates the presence or absence of true seizure activity. The noisy training label~$y_n$ (often provided by clinicians) is also a binary random variable, but it is influenced by both the observed EEG data~$x$ and the true seizure label~$y_c$. During training, $y_n$ is available for the model to use, while $y_c$ remains unknown. During testing, the model relies solely on the learned parameters without access to any annotations. In simulated experiments, we set aside $y_c$ to evaluate the model, using $y_n$ for real-world datasets like TUH and CHB-MIT, where clean annotations are unavailable. Finally, the graphical model assumes an instantaneous relationship between variables, meaning the noise corruption of a particular label does not depend on past or future time points. 

\subsubsection*{Overview}

Our high-level strategy is to recognize that the training labels of seizure activity, derived from clinical review of the EEG, may be unreliable when the data contains significant noise or artifacts. Thus, we develop a statistical framework that models the mismatch between the given labels~$y_n$ and the (unseen) ground truth labels~$y_c$, which we refer to as ``clean" labels. Based on this framework, we learn the function $f_\theta(\cdot)$, parameterized by a deep neural network parameter~$\theta$, infer the clean label posterior distribution~$p(y_c|x)$ given the EEG data~$x$ in a weakly supervised manner. This strategy allows us to predict true seizure activity given noisy annotations. %during training. 
Note that $\theta$ can refer to any deep network that incorporates dropout, making BUNDL model-agnostic. {All mathematical notations are described in Table~\ref{tab:variables}}

 \begin{table*}[!ht]
        \begin{adjustwidth}{-1.in}{0in} % Comment out/remove adjustwidth environment 
        \centering
        \caption{\textbf{Description of all mathematical notations in the mansucript}}
        \label{tab:variables}
        \begin{tabular}{|l|l|l|l|}
        \hline
        Description      & Variable           & Description         & Variable              \\ \hline
        \begin{tabular}[c]{@{}l@{}}Clean label \end{tabular}    & $y_c$     & Clean label Bernoulli parameter   & $p_{y_c}$ \\
        \begin{tabular}[c]{@{}l@{}}Noisy label \end{tabular} & $y_n$  & Noisy label Bernoulli parameter & $p_{y_n}$   \\ 
        \begin{tabular}[c]{@{}l@{}} Input EEG \\ \end{tabular}  & $x$        & Optimal weights after finetuning  & $\theta^*$             \\ 
        \begin{tabular}[c]{@{}l@{}}Label unreliability measure \\ \end{tabular} & $z_{xy_n}$  & Pretrained deep network weights     & $\theta_{pre}$        \\ 
        \begin{tabular}[c]{@{}l@{}} Deep network function \\ \end{tabular}   & $f_\theta(\cdot)$   & Learning rate   & $\alpha$ \\ 
        \begin{tabular}[c]{@{}l@{}} Single model output (n - index) \\ \end{tabular}  & ${f^n} = f_{\theta}(x)$    & Number of multiple outputs     & N                     \\ 
        \begin{tabular}[c]{@{}l@{}} Average prediction \\  \end{tabular}         & $\overline{f_\theta(x)} = \frac{1}{N}\sum_{n=1}^{N} {f}^n$ & Normalizing parameter        & $C = \log2$           \\ 
        \begin{tabular}[c]{@{}l@{}} Bernoulli distribution \\   \end{tabular}   & $\mathcal{B}(y_c; p_{y_c})$   & Maximum number of epochs   & M                     \\ 
        \begin{tabular}[c]{@{}l@{}} Loss criteria \\ \end{tabular}     & $\mathcal{L}$        & KL divergence            & $D_{KL}(\cdot)$       \\ \hline
        \end{tabular}
        \end{adjustwidth}
    \end{table*}
    
Formally, given the training EEG data and seizure versus baseline labels~$(x,y_n)$, we define the noisy label posterior distribution as $p(y_n|x)=\mathcal{B}(y_n; p_{y_n})$, where $\mathcal{B}(\cdot)$ denotes a Bernoulli distribution, and the parameter $p_{y_n}$ denotes the probability of seizure activity. In our case, $p_{y_n}$ equals the binary label of seizure activity provided by the clinician, but it could alternatively equal a continuous measure of clinician confidence, should that information be available. {We account for the mismatch between the unknown ``clean" label $y_c$ and the observed and potentially ``noisy" label $y_n$ for the given EEG window $x$ as follows}

\begin{eqnarray} \label{eqn:cleanpost}
    p(y_c|x, y_n) = \mathcal{B}(y_c;  p_{y_n}(1-z_{xy_n}) + \overline{f_\theta(x)}z_{xy_n} ),
\end{eqnarray}
{  where $z_{xy_n}$ quantifies the unreliability of the input data and given label ($x$, $y_n$), and $\overline{f_\theta(x)}$ is the neural network prediction for the EEG window using parameters of the current epoch. For robustness against noise, $\overline{f_\theta(x)}$ is quantified as the average of multiple forward passes to the model. The uncertainty measure $z_{xy_n}$ can be estimated based on the consistency of the neural network outputs $\overline{f_\theta(x)}$. Our formulation is motivated by prior work in the deep learning literature, which suggests that the softmax probability outputs produced by neural networks contain information about label noise/uncertainty after just a few epochs of training with a generic loss~\cite{algan2020label,goel2021robustness,li2021improved}. This observation can be attributed to a ``memorization effect", as models tend to fit the easier samples with clean/unambiguous labels first~\cite{liu2020early,wei2023mitigating}. In contrast, samples corrupted by label noise are more challenging and are fitted in later training stages; thus, models will have lower confidence in these sample predictions early on. Intuitively, when   $z_{xy_n}$ is low, the model will trust the given labels $p_{y_n}$ provided by clinicians. In contrast, when the  $z_{xy_n}$ is high, the model hinges towards its previous predictions, $\overline{f_\theta(x)}$ and ignores the given labels.} 

We multiply Eq.~(\ref{eqn:cleanpost}) with $p(\hat{y}|x)=\mathcal{B}(\hat{y}; p_{g})$ to obtain the joint distribution $p(y, \hat{y}|x)$. Finally, we marginalize $\hat{y}$ from the joint to obtain the clean label posterior $p(y|x) = \mathcal{B}(y; p_{yc})$ as follows:

\begin{eqnarray} \nonumber
    %p(y_c|x) &= \sum_{y_n=0}^{1} p(y_c|x, y_n) p(y_n|x) \\ \nonumber
    p(y_c|x) &= \underbrace{\sum_{y_n=0}^{1} \mathcal{B}(y_c;  p_{y_n}(1-z_{xy_n}) + \overline{f_\theta(x)}z_{xy_n} ) \cdot \mathcal{B}(y_n; p_{y_n})}_{\mathcal{B}(y_c; p_{y_c})},
\end{eqnarray}
where the clean Bernoulli parameter~$p_{y_c}$ can be computed as
\begin{align} \nonumber
    p_{y_c} &= p_{y_n} (z_{x1} \cdot \overline{f_\theta(x)} + p_{y_n}(1-z_{x1})) \\
    &+ (1-p_{y_n})(z_{x0} \cdot \overline{f_\theta(x)} + p_{y_n}(1-z_{x0})) \label{eqn:cleanparam} 
\end{align}

During training, we minimize the KL divergence between the Bernoulli distribution $\mathcal{B}(y_c; p_{y_c})$  and the  network prediction $Q(y_c|x ; \theta) = \mathcal{B}(y_c; f_\theta(x))$ as follows: 
\begin{align}
    \theta^* &= \arg\min_\theta D_{KL}(P||Q)  \nonumber \\
     &= \arg\max_\theta \Big(p_{y_c} \cdot log(f_\theta(x)) + (1-p_{y_c}) \cdot log(1-f_\theta(x)) \Big) \label{eqn:min}
\end{align}
The optimal deep network parameters~$\theta^*$ are used  directly to predict  clean labels. Thus, BUNDL aligns the network predictions with the  underlying clean label distribution.% of the clean labels.

\subsubsection*{{Approximating label mismatch with $z_{xy_n}$}}

{BUNDL relies on the unreliability measure $z_{xy_n}$ to determine if the clinician-provided labels should be trusted during training. Effectively, when $z_{xy_n}$ is high, the labels are considered unreliable. Since we do not have ground truth information about~$z_{xy_n}$, we must either fix it \textit{a priori}, as suggested by~\cite{huang2020self}, or estimate it from the network's outputs. For example, the work of \cite{li2021improved} used the loss function value in each epoch to identify incorrect labels (e.g., high loss values) and correct them in the subsequent training epochs. This procedure is based on the idea that models learn cleanly labeled samples faster than noisy ones, which are harder to learn from~\cite{liu2020early,wei2023mitigating,li2021improved}}. 

{Prior work also suggests that label noise is reflected in the softmax probabilities produced by the neural network due to memorization effect i.e, they fit easier on samples with clean labels~\cite{algan2020label,goel2021robustness}, particularly after the early stages of training. Building on this observation, we hypothesize that a neural network will be uncertain about its prediction on a challenging sample, which will be reflected in the \textit{variability of the model output}. Higher variability in the model output is equivalent to having flatter output probability distribution across classes, i.e., higher entropy. Therefore,  we propose to use Monte Carlo Dropout (MCD) to estimate average entropy with respect to the input data across multiple forward passes through the model. Importantly, MCD allows us to estimate $z_{xy_n}$ from the existing deep network architecture without adding new parameters~\cite{gal2016dropout}. Our approach leverages MCD during both training and inference to capture the variability in predictions, thus providing insights into the confidence and the reliability of the model outputs. While MCD is often used to quantify epistemic uncertainty, it is simultaneously true that entropy at the output is higher for class-ambiguous labels due to memorization effect, leading to gradient inconsistencies during training~\cite{algan2020label}. As shown in Fig~\ref{train}, we use dropout to make multiple predictions and estimate an empirical distribution for $p(y_c|x)$. We then compute  $z_{xy_n}$ as the self-entropy across N = 20 MC samples as follows:}
    \begin{eqnarray} \label{eqn:unc}
    z_{xy_n} = -\frac{1}{CN} \sum_{n=1}^N \left[ {f}^n \log {f}^n + \left(1-{f}^n\right) \log \left(1-{f}^n\right) \right]
    \end{eqnarray} 

{The division by $C = \log(2)$ scales all values to lie between 0 and 1 and the variable ${f}^n$ corresponds to multiple predictions from the model using the randomization provided by MCD.Our use of MC samples for weak supervision in Eq.~(\ref{eqn:unc}) is akin to having multiple seizure labels from different clinicians. This strategy mitigates the influence of faulty predictions on the learning process.}

{The MCD-based entropy used to quantify the uncertainty~$z_{xy_n}$ in Eq.~(\ref{eqn:unc}) is bounded and is considered robust in noisy label conditions. Specifically, prior work has found that dropout makes the network sparser and acts as a regularizer to prevent models from being overly confident in its predictions~\cite{srivastava2014dropout}. Thus, high~$z_{xy_n}$ encourages the label transition probability in Eq.~(\ref{eqn:unc}) to be larger, as there is that is a higher chance of the given labels being incorrect. Alternate methods for uncertainty quantification include (i)~ensemble-based entropy estimation which generates multiple samples from multiple models trained with different initialization~\cite{srivastava2014dropout,borovac2022ensemble}; (ii)~test-time augmentation (TTA) which computes the entropy over the outputs generated from randomly perturbed input samples~\cite{10340330,smith2025uncertainty}; (iii)~the loss value between the predicted and provided labels, which is inspired from prior literature suggesting that label noise proves to be harder for models to fit~\cite{li2021improved}; and (iv) a fixed level of 0.9 mismatch in labels suggested by~\cite{huang2020self}. Using simulated data, we compare these for alternative uncertainty quantification methods with MCD. To compare behavior, we compute the average $z_{xy_n}$ separately for samples with correct labels and with incorrect labels. As seen in Table~\ref{tab:uqrange}, MCD, ensemble, TTA yield statistically different values for $z_{xy_n}$ and thus can be used to detect input-dependent label noise. The constant method makes no effort in distinguishing which samples with label noise by definition. Surprisingly, loss as a label unreliability measure also had a very small range for $z_{xy_n}$, likely due to the unbounded nature of the cross-entropy loss function,in which large values suppress variations during normalization.}

    \begin{table*}[!ht]
    %\color{black}
    \begin{adjustwidth}{-0.75in}{0in} % Comment out/remove adjustwidth environment 
    \centering
    \caption{{\textbf{Range of  $z_{xy_n}$ using different methods in our ablations on the simulated dataset and DeepSOZ architecture.} }}
    \label{tab:uqrange}
    \begin{tabular}{|l|l|l|l|l|l|}
    
    \hline
    {Simulated data} & {Constant~\cite{huang2020self}}    & {Loss \cite{li2021improved}}       & {TTA\cite{10340330,smith2025uncertainty}}         & {Ensemble\cite{srivastava2014dropout,borovac2022calibration}}    & {MCD\cite{gal2016dropout}}         \\ \hline
    {Correct}           & {0.900 $\pm$0.00}  & {0.001 $\pm$0.0}  & {0.312 $\pm$0.240} & {0.529 $\pm$0.165} & {0.136 $\pm$0.173} \\
    {Incorrect}         & {  0.900 $\pm$0.001} & {0.001 $\pm$0.01} & {0.456 $\pm$0.379} & {0.667 $\pm$0.037} & {0.326 $\pm$0.339} \\
    {p-value}           & {$\sim$1.0}   & {0.848}      & {0.001}       & {$\sim$0.0}   & {$\sim$0.0}  \\ \hline
    \end{tabular}
    \begin{flushleft}Mean and standard deviation per input samples with correct labels and incorrect labels (mismatch in noisy and true annotations) across cross validation folds are reported. p-value is reported from t-test under the null hypothesis that the computed $z_{xy_n}$  is same within sampled with correct and incorrect labels.\end{flushleft}
    \end{adjustwidth}
    \end{table*}

Eq.~(\ref{eqn:unc}) allows us to incorporate \textit{a priori} knowledge about the expected label noise. For example, if we expect both seizure and non-seizure labels to have the same level of uncertainty, then we would set $z_{x0} = z_{x1}$, as currently shown in Eq.~(\ref{eqn:unc}). However, it is often the case that clinicians over-segment the seizures to avoid missing problematic time intervals~\cite{amin2019role}. To accommodate this tendency, we fix $z_{x0} = 0.001$, a small quantity, while $z_{x1}$ is derived from Eq.~(\ref{eqn:unc}). Conversely, if seizures are known to be under-segmented, we could fix~$z_{x1}$ at a small value while estimating $z_{x0}$ during training. Hence, the user can adapt BUNDL to different application domains. %based on the presumed noise characteristics.

{While the unreliability $z_{xy_n}$ computed via MCD is influenced by epistemic uncertainty, the presence of label ambiguity also pushes the output entropy to be higher. Thus, MCD is a reasonable proxy for aleatoric influences related to the label noise itself~\cite{algan2020label}. Similar to how clinician fatigue or limited experience can lead to mislabeling \cite{halford2013standardized,halford2015inter}, challenging EEG samples are more likely to be annotated incorrectly, motivating the use of model uncertainty to flag potentially noisy labels. In order to obtain good estimates for  $z_{xy_n}$ and $\overline{f_\theta(x)}$,} we pretrain the base network using a cross-entropy loss on EEG samples that lie squarely within baseline and seizure regions~\cite{chen2019understanding}. In this case, we are more certain that the clinician annotated labels are correct, as ambiguity tends to be highest around the seizure onset and offset times. This pretraining increases model confidence and encourages the uncertainty computed in Eq.~(\ref{eqn:unc}) to reflect the desired {label noise.}  Hence, we believe that our subsequent training using BUNDL we can teach the model to recognize faulty labels and learn true seizure characteristics.

%\medskip
\subsubsection*{Training algorithms}

\captionsetup[algorithm]{font=small, skip=10pt}

Following the pretraining phase, we utilize the steps described in  Algorithm~\ref{alg:train} and Fig \ref{train} (middle)  to train  deep networks  with BUNDL. Each training epoch includes making a prediction, computing the uncertainty using the MC dropout procedure in Algorithm~\ref{alg:cap}, and a gradient update to tune the deep network weights without additional parameters. In addition, the efficiency of Algorithm~\ref{alg:cap} (MC samples) can be improved using parallel  processing.

    \begin{algorithm}[!h]
    \caption{BUNDL Training Framework}
    \label{alg:train}
    \begin{algorithmic}[1] % [1] for line numbers
    \REQUIRE Learning rate $\alpha$, Max epochs $M$, $\{x, p_{y_n}\}$ batched input of EEG-noisy label pairs, and optimizer
    \STATE $f_\theta(\cdot)$ given deep network architecture, $\mathcal{L}$ is the KL divergence loss from equation \ref{eqn:min}
    \STATE $z_{xy_n}$ label unreliability level where $y_n=0$ (non-seizure) or $y_n=1$(seizure) 
    \STATE $\theta \gets \theta_{pre}$ \COMMENT{if pretrained model given}
    \STATE $m \gets 1$
    \WHILE{$m \leq M$}
        \FOR{$(x, p_{y_n}) \in$ train data}
            \STATE Compute $\overline{f_\theta(x)}$, and $z_{xy_n}$ using Algorithm \ref{alg:cap}
            \STATE $p_{y_c} \gets p_{y_n} (z_{x1} \cdot \overline{f_\theta(x)} + (1-z_{x1})p_{y_n}) + (1-p_{y_n})(z_{x0} \cdot \overline{f_\theta(x)} + (1-z_{x0})p_{y_n}) $
            \STATE $\mathcal{L} \gets -p_{y_c} \cdot \log(f_\theta(x)) - (1 - p_{y_c}) \cdot \log(1 - f_\theta(x))$
            \STATE $\theta \gets \theta - \alpha \nabla \mathcal{L}$ \COMMENT{Update with optimizer}
        \ENDFOR
        \STATE $m \gets m + 1$
    \ENDWHILE
    \end{algorithmic}
    \end{algorithm}
    
    \begin{algorithm}[!h]
    \caption{BUNDL Uncertainty Computation}
    \label{alg:cap}
    \begin{algorithmic}[1] % [1] for line numbers
    \REQUIRE  Deep network $f_{\theta}(\cdot)$, input EEG $x$ from current epoch, and  MC samples $N$
    \ENSURE No gradient computation
    \STATE Normalizing constant $C \gets ln2$,  Counter $n \gets 1$
    \STATE Outputs $f \gets \mathbf{0} \in \mathbb{R}^{N}$
    \WHILE{$n \leq N$}
        \STATE ${f^n} \gets f_{\theta}(x)$ \COMMENT{Network output with dropout}
        \STATE $n \gets n + 1$
    \ENDWHILE
    \STATE Average output $\overline{f_\theta(x)} \gets \frac{1}{N} \sum_{n=1}^{N} {f}^n$
    \STATE Label uncertainty $z_{xy_n} \gets \frac{-1}{C N} \sum_{n=1}^{N} \left({f}^n \log {f}^n + (1 - {f}^n) \log (1 - {f}^n)\right)$
    \end{algorithmic}
    \end{algorithm}

\subsection*{Baseline comparison methods} 

We compare BUNDL with two state-of-the-art approaches for handling noisy training labels. The first approach is Self-Adaptive Learning (SelfAdapt)~\cite{huang2020self}, which uses a weighted sum of given labels and past predictions as a soft-pruning to re-weight and/or leave out noisy samples in loss computation. The second approach is a Noisy Adaptation Layer (NAL)~\cite{goldberger2016training}, which uses an EM-like algorithm to jointly estimate both clean and noisy label posteriors. We use the ``complex model" variant introduced by the authors, as it is better suited for seizure detection. Finally, we include a the traditional cross-entropy loss (CEL) as a baseline, which does not account for label noise and assumes that the provided labels are accurate.

\subsection*{Implementation and evaluation}

As noted, BUNDL can be used to address noisy labels by training any deep network. To underscore this, we apply BUNDL to three state-of-the-art seizure detection networks.

\begin{enumerate}
    \item \textbf{DeepSOZ:} Our recently-introduced DeepSOZ model~\cite{m2023deepsoz} consists of a transformer that extracts both global features across the EEG channels and channel-wise features for each time window. The global features are processed by a long short-term memory (LSTM) network to predict the occurrence of seizure activity (i.e., seizure detection). Simultaneously, the channel-wise features are passed through a pooling layer to aggregate information across time  enabling DeepSOZ to identify the channels  associated with the onset of seizures (i.e., seizure onset zone localization).

    \item \textbf{TGCN:} The TGCN model introduced by~\cite{covert2019temporal} also captures both channel-wise and temporal patterns in EEG data via eight layers of spatio-temporal convolutions (STC). The architecture leverages 1D convolutions that aggregate the EEG data in the neighborhood of each channel. This approach allows TGCN to model the propagation of seizure activity across EEG channels and time. Following the STC layers, three linear layers are used to detect seizures from the extracted features. % seizure detection).

    \item \textbf{CNN:} The CNN model by~\cite{craley2021automated} combines convolutional (1D) and recurrent layers to process EEG signals for seizure detection. %The CNN component uses 1D convolutions to capture broad temporal features across all EEG channels for each time window. 
    Convolutions followed by Batch Normalization are applied to enhance training. The output is then fed into a bidirectional LSTM layer, which captures temporal dependencies for  detection.
\end{enumerate}

{ Finally, we implement the Hybrid Vision Transformer architecture with Data Uncertainty Learning (HViT-DUL) recently presented in~\cite{deng2023eeg}. While HViT-DUL is designed for noisy EEG data, rather than label uncertainty, its architecture is inspired by Bayesian neural networks, and provides an interesting benchmark for BUNDL. We re-implement the originally proposed model with slight modification the architecture to handle the EEG electrode setup and preprocessing of our datasets, while following the training algorithm proposed in the original manuscript. For application on our dataset, the kernel size of the input layer of HViT was modified to fit inputs from 19 channels of 1 second EEG in TUH and Siena datasets, and 18 channels of 1 second EEG in CHB-MIT, all preprocessed at 200Hz.}

{Each of the networks is trained with BUNDL, the noisy label comparison methods (SelfAdapt \& NAL), and the baseline CEL separately. For BUNDL, we initialize the model trained on CEL loss with samples well within the seizure and non-seizure class trained. We train the models in a nested 10 fold cross-validation setup, repeated 5 times to have 50 models trained per method. We use Adam optimizer using PyTorch 1.10 for 30 epochs in the stimulated and TUH datasets and for 50 epochs in CHB dataset for pretraining and finetuning.  NAL has two stages of pretraining for the base model and for the simple model, and finally finetuning of noise adaptation layers each for 30 epochs in TUH and simulated datasets 50 epochs in the CHB dataset. HViT-DUL\cite{deng2023eeg} network is trained the loss function provided in their original publication with same number of maximum epochs as BUNDL. Early stopping tolerance of 10 epochs with no improvement in validation loss is used in all methods. The learning rates for all methods are  chosen separately in the range of [$0.01-10^{-6}$]  by assessing the loss curves in the cross-validation setup for smooth descent and lowest validation loss before testing.All networks have dropout percentage at 20\% and 20 MC samples in the algorithm. We also use a small tolerance of 0.001 for $p_{y_n}$ to ensure numerical stability. Our code is available on \href{https://github.com/deeksha-ms/BUNDL}{Github.}}

We evaluate the seizure detection performance at two levels. At the window level, we report the area under the receiver operating characteristic curve (AUROC) and the area under the precision-recall curve (AUPRC), which summarize the performance on the individual EEG time windows. At the seizure level, we assess performance across the entire recording and report the percentage of seizures detected (sensitivity), the false positive rate in minutes per hour (FPR), and the latency in detecting the seizure onset. The detection threshold for when a predicted probability is considered to be a seizure, is a hyperparameter chosen through nested cross-validation. Specifically, it is selected from the range [0.1, 0.8] to maximize sensitivity while ensuring a false positive rate of less than 3 minutes/hour on the nested validation dataset.

\subsection*{SOZ localization analysis}

While most deep learning methods focus on seizure detection, determining region in which the seizure originates, also known as seizure onset zone (SOZ) localization, is arguably the more important clinical task~\cite{rosenow2001presurgical}. When the SOZ can be narrowed to a discrete region in the brain,  surgical resection of this area can provide the highest rate of seizure freedom~\cite{olmi2019controlling, rosenow2001presurgical}. DeepSOZ is designed to perform both seizure detection and SOZ localization. We conjecture that the SOZ localization performance is closely tied to its seizure detection accuracy due to the link between the global and channel-wise features. In this study, we analyze the impact of using BUNDL  on SOZ localization, as compared to the conventional CEL objective function. in repeated 10-fold cross-validation setup similar to detection, repeated 5 times, to select the learning rate for training with Adam optimizer in PyTorch 1.10 and to evaluate performance.

\subsection*{EEG datasets}
{We use four datasets in our experiments as summarized in Table \ref{tab:demo_table}: One is a  simulated dataset designed to emulate various noisy label settings, and the other three are real-world publicly available datasets~\cite{shah2018temple,guttag2010chb,detti2020siena}. All real world-datasets have been fully anonymized prior to release under the ethics requirements of each institution. For example, the TUH dataset was reportedly collected in accordance with the Declaration of Helsinki ensuring informed consent and with the full approval of the Temple University Hospital IRB~\cite{shah2018temple}. Based on these factors, the Charles River Campus IRB at Boston University has waived an ethics review of this study.}.  

\begin{table*}[!ht]
\begin{adjustwidth}{-1.8in}{0in} % Comment out/remove adjustwidth environment if table fits in text column.
    \centering
    \caption{\textbf{Description of siezure characteristics in simulated,  TUH, CHB-MIT, {and Siena datasets}.}}
    \begin{tabular}{|c||c|c|c|c|}
    %\hline
    \hline
    & Simulated dataset &  TUH dataset & CHB-MIT dataset & {Siena dataset}\\
    \hline   
    \hline
    Number of patients & 120 &120& 23 & {14}\\ 
    %\hline   
    Total Number of seizures &631 & 642 & 181 & {47}\\
    %\hline   
    Average seizures per patient &  5.3$\pm$2.69 &14.7$\pm$25.2 & 7.9$\pm$6.0 & {3.35$\pm$2.31}\\
    %\hline   
    Min/Max seizures per patient & 1/10 &1/152  & 3/27 & {1/10}\\
    %\hline   
    Average EEG duration per patient & 52.8$\pm$26.97 min & 79.8$\pm$135 min  & 370$\pm$182.4 & {269.7$\pm$240.2 min} \\ 
    %\hline   
    Average seizure duration  &   170.5$\pm$120.4 sec  &88.0$\pm$123.5 sec & 63.6$\pm$74.0 sec & {68.1$\pm$26.32 sec}\\
    %\hline   
    Min/Max seizure duration & 29/491 sec &7.5/1121 sec & 7/753 sec & {33.8/122 sec}\\
    \hline   
    \end{tabular}
    \label{tab:demo_table}
   % \begin{flushleft} Table notes \end{flushleft}
\end{adjustwidth}
\end{table*}

\subsubsection*{Simulated data}
\begin{figure}[!h]
        \centering
        \includegraphics[width=0.75\linewidth]{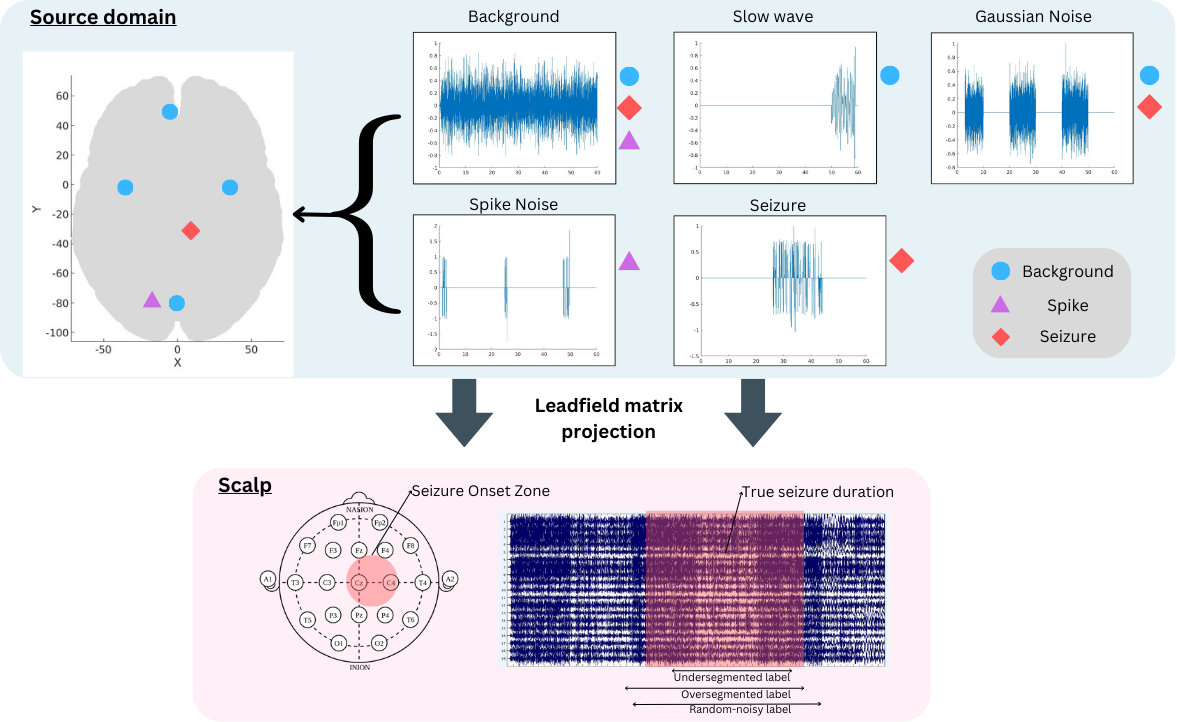}
        \caption{{\textbf{Simulated EEG and label noise generation pipeline using SEREEGA~\cite{krol2018sereega}.} The top half shows components in source domain including source positions and five types of signals assigned to corresponding source indicating with following icons: background is shown in blue circles, the spike noise in purple triangle, and seizure source in red diamond. The background sources are fixed in position while the latter two is randomly assigned per simulated participant. The bottom half shows the scalp electrode plot of 10-20 montage and an example EEG with true seizure and noisy annotations marked.}}
        \label{fig:sim_gen}
    \end{figure}

% From the 32-channel New York Head (ICBM-NY) leadfield matrix, we select 10-20 channels per recording. 
We use the SEREEGA framework~\cite{krol2018sereega} in MATLAB to simulate EEG data for 120 subjects {as depicted in Fig~\ref{fig:sim_gen}}. SEREEGA uses a source-scalp model based on the lead-field matrix and randomized source locations for each subject~\cite{huang2016new}, to emulate real-world signal perturbations associated with seizures. We randomly generate between $1-10$ EEG recordings for each subject, with each recording lasting 10~minutes. All recordings are sampled in the 10-20 EEG montage at 200 Hz. Seizure onset and duration are randomly determined for each recording, with seizure durations ranging between $29-491$ seconds (average 170± 120.4 seconds). Outside these seizure intervals, the EEG signals include a background of alpha and beta waves, and additional  components as described below.
\begin{itemize}
    \item \textbf{Seizure Characteristics:} Each subject is assigned a single seizure source characterized by spikes, sharp waves, and polyspike bursts within the $2.5-4$~Hz range. This source is randomly activated during the seizure period and remains fixed in location for each subject. 
    \item \textbf{High-Frequency Noise:} High-amplitude one-sided spikes of 6-14Hz, lasting less than 10 seconds, are introduced at random times and regions in each signal.  

    \item \textbf{Low-Frequency Noise:} One source in the posterior and anterior regions of the brain is assigned to generate slow waves ($1-3$~Hz) , lasting $5-60$~seconds. %The duration of these waves is randomly set between  for each subject.
    
    \item \textbf{Gaussian Noise:} Random bursts of Gaussian noise are applied to all sources. Gaussian noise is also added 30 seconds around seizure onset. Noise is generated at an Signal-to-Noise ratio (SNR)  of $-1.6-0.92$ dB. Effect of EEG SNR level is discussed in  \nameref{S1_Appendix:snr} and \nameref{S1_snr}.
\end{itemize}

{After generating the 10-20 EEG recordings using SEREEGA, we segment the data into 1-second non-overlapping windows and assign a ``clean" label (seizure vs. baseline) to each window using the ground truth seizure timing information. Next, we introduce label noise at levels comparable to the real-world inaccuracies that might be observed during clinical review. Namely, in a study with 9 expert annotators and 991 EEG recordings~\cite{jing2020interrater}, the inter-annotator agreement in identifying epileptiform activity was 67.0\%-77.8\% for each event, suggesting that on average 20-30\% label noise can be expected in single annotator. We generate five cases of label noise as follows:}
    \begin{enumerate}
        \item {Symmetric label noise: Given the true seizure time annotations [$t_{start}$, $t_{end}$] from the EEG generation process, we create a noisy annotation by randomly perturbing the seizure onset to lie in the interval [$t_{start}$-30, $t_{start}$+30] and the seizure offset to lie in the interval [$t_{end}$-30, $t_{end}$+30] with at least 5 seconds of overlap with the original seizure interval. We add Gaussian noise to the EEG data within the perturbed seizure interval to blur the exact seizure timing. At a high level, this noise configuration randomly changes the length and position of the seizure interval. We verified that the resulting label noise was empirically within the 0-25\% range. The corruption is also symmetric, as both seizure and baseline labels are corrupted at random with no class-specific knowledge for BUNDL to incorporate.}  
        
        \item {Over-segmentation of seizures at 10\% and 30\%: In these cases, the noisy seizure labels are created to be longer than the ground truth clean labels. This process is implemented by randomly selecting the seizure onset and offset times, such that the seizure duration increases by 10\% (case 1) or 30\% (case 2). If the over-segmentation exceeds the 10-minute recording length, then the noisy seizure labels are truncated to 1 minute outside the true seizure interval. over-segmentation is the most common type label noise observed in seizure annotation by clinicians~\cite{halford2013standardized, halford2015inter,jing2020interrater}, as missing seizure activity is more detrimental than being overly generous in the seizure onset and offset. From a machine learning standpoint, this over-segmentation poses as a one-sided challenge when training deep neural networks.}

        \item {Under-segmentation of seizures at 10\% and 30\%: As a complement to the over-segmentation, we use a similar procedure to \textit{reduce} the seizure duration by 10\% (case 1) and 30\% (case 2), while ensuring that the seizure lasts $\geq$29 seconds. Seizures that are too short to be reduced by the required percentage are capped at 29 seconds. We expect under-segmentation to be rare in real-world data, as clinical review is biased against missing seizure activity.}
        \end{enumerate}

{In total, we conduct experiments based on the five label noise cases described above. For each case, we simulate EEG data with three SNR levels, yielding a total of 15 configurations. We use only the noisy labels for model training and evaluate performance on the unseen clean labels.}

%\medskip \noindent
\subsubsection*{TUH dataset}
We collect 120 subjects with focal seizure onset from the publicly available Temple University Hospital (TUH) corpus \cite{shah2018temple}. The anonymized dataset was first accessed on 22 May 2022. There are 55 male and 65 female subjects within the age range of 19-91 years (55.2$\pm$16.6). Each seizure recording includes average-referenced EEG data from 10-20 montages, along with clinician-labeled seizure intervals and onset channels. We resample the EEG signals to 200~Hz for consistency with the other datasets. We apply minimal preprocessing that includes a bandpass filter ($1.6-30$~Hz) and clipping the signal at two standard deviations from the mean to eliminate high-intensity artifacts in EEG. The EEG signals are standardized to have a zero mean and unit variance using statistics derived from each patient to avoid any information leakage. The recordings are cropped to 10 minutes around the seizure events with an even distribution of  onset times and  divided into one-second, non-overlapping windows. 

For the downstream analysis of SOZ localization, we use the clinician notes to extract the onset channels for each patient in  TUH. This results in 72 seizures originating from the temporal lobe and 48 from extra-temporal lobes. %There are 59 seizures in the left hemisphere and 61 in the right. 
While the localization data is not used during detection, it plays a crucial role in epilepsy treatment. Thus, we evaluate how accounting for label noise during detection  impacts the  localization performance.

%\medskip \noindent
\subsubsection*{CHB-MIT dataset}

The CHB-MIT public dataset provides EEG recordings from 23 pediatric subjects (5 male, 17 female) aged 1.5 to 22 years~\cite{guttag2010chb,shoeb2009application,goldberger2000physiobank}. The fully anonymized was first accessed on 13 February 2024. We use 18 channels from the bipolar-referenced 10-20 montage and resample the data from 256 Hz to 200 Hz. Each 60-minute segment contains at least one seizure. %, and the dataset includes various seizure types. 
Patients have between 3 and 27 seizures, averaging 7.9$\pm$6.0 minutes in duration, with some exhibiting multiple seizures within a single recording. We apply similar preprocessing steps as for the TUH dataset, which include filtering, artifact removal,  normalization, and segmenting into 1-second non-overlapping time windows, thus ensuring uniformity across datasets.% and enhances the robustness of our seizure detection models.

\subsubsection*{Siena dataset}

{The publicly available Siena dataset includes EEG recordings from 14 adults (8 male, 6 female) aged 20–58 years \cite{detti2020siena,detti2020eeg,goldberger2000physiobank}. The dataset was first accessed on 11 December 2024. We use 19 channels from the average-referenced 10–20 montage and resample the original 256 Hz data to 200 Hz, in order to remain consistent with TUH. Each 10-minute segment contains at least one seizure, with some segments containing multiple seizure events. All patients have between 1-10 seizures in total. Preprocessing followed the same steps as the TUH dataset, which included filtering, artifact removal, normalization, and segmentation into 1-second non-overlapping windows.}
% Results and Discussion can be combined.
\section*{Results}

\subsection*{Simulated experiments}
We perform a comprehensive evaluation of BUNDL and the baseline methods by applying them to three deep network models under a variety of lable noise settings. As we have access to the ground-truth seizure labels for simulated data, we use  noisy labels only for model training  (as in real-world conditions),
and  evaluate against true  labels.

\subsubsection*{Convergence analysis}

\begin{figure}[!h]
    \centering    
   \includegraphics[width=0.75\columnwidth]{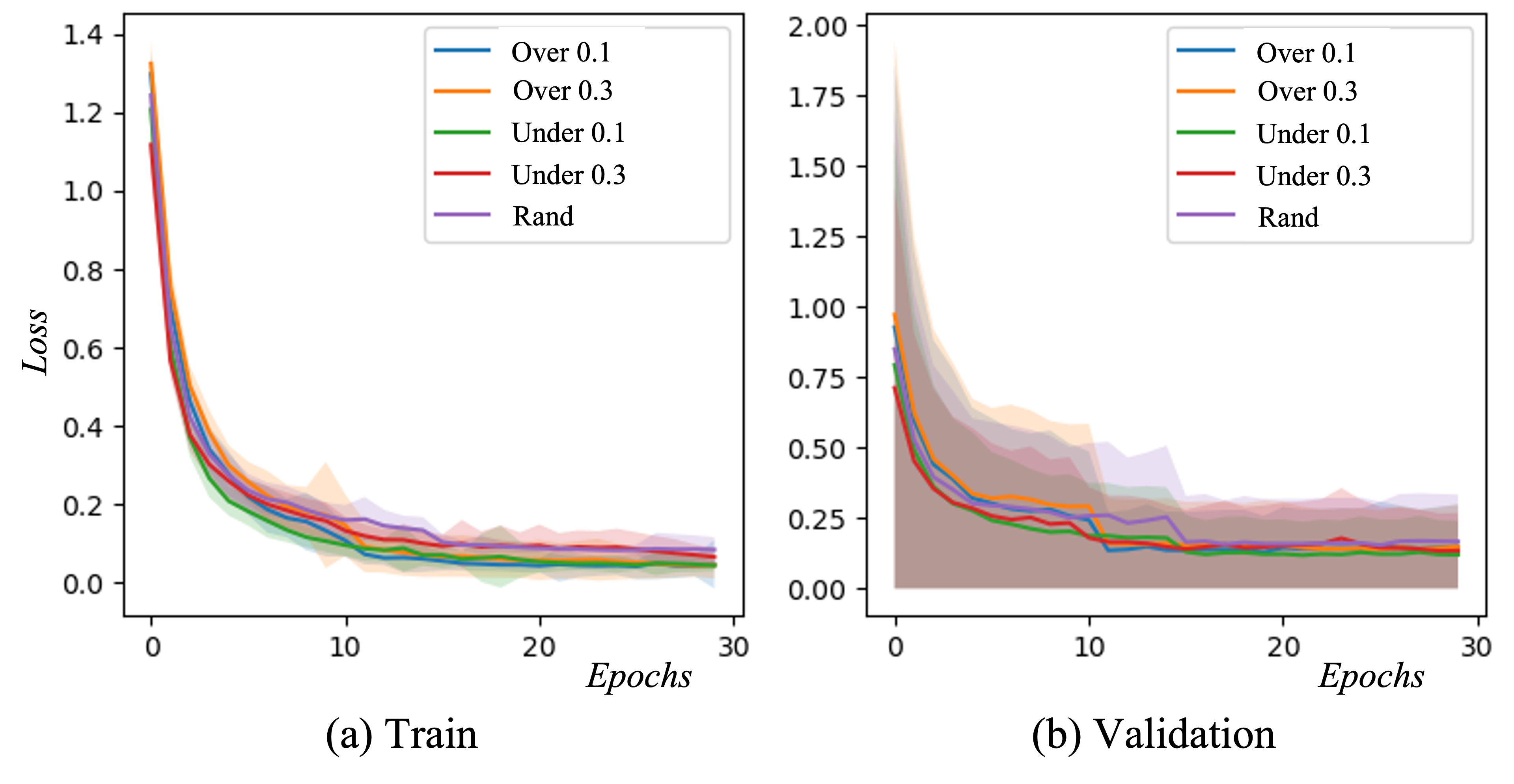}
    \caption{\textbf{Loss curves}  (a) from training and (b) from validation folds of all simulated experiments of training CNN with BUNDL. }
    \label{loss}
\end{figure}

All noisy label methods (BUNDL, SelfAdapt, NAL, CEL) consistently converged across all experimental settings. An example of training the CNN with BUNDL is shown in Fig~\ref{loss}. Empirically, we observe that BUNDL converges the fastest when initialized from pretrained models. In contrast, Self-Adapt was slower to train and required additional memory to store past predictions -- an overhead avoided in BUNDL through parallel batch processing of MC samples. NAL was the slowest, as its more complex variant required two separate finetuning stages to capture data-dependent label transitions. These differences make BUNDL particularly appealing for real-world deployment, where faster training and lower memory usage are critical for scalability and adaptability.

\subsubsection*{Performance analysis}

\begin{figure}[!h]
    %\centering    
   \includegraphics[width=0.95\linewidth]{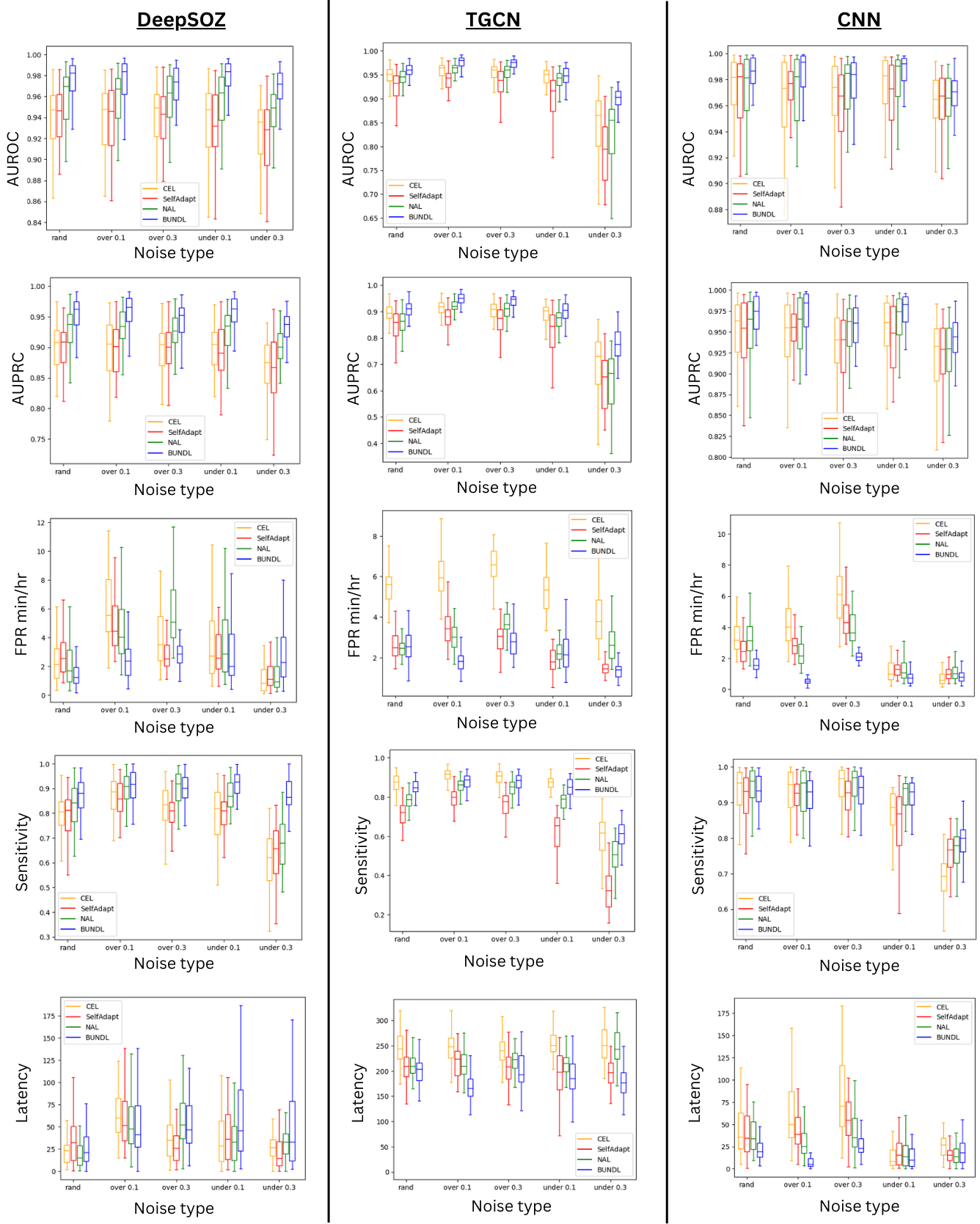}
    \caption{\textbf{Box plots of performance metrics of three models trained using various strategies across five label noise types.} Results are organized by deep network across columns - (left) DeepSOZ, (middle) TGCN, and (right) CNN - and by evaluation metric across rows: AUROC, AUPRC, FPR (min/hour), Sensitivity, and Latency (seconds). Each subfigure presents performance from BUNDL alongside three baseline comparisons at various label noise cases of rand-symmetric noise, over 0.1-over-segmented at 10\%, over 0.3-oversegmetned at 30\%, unner 0.1-under-segmented at 10\%, under 0.3-undersegmetned at 30\%
    }
    \label{labelnoise}
\end{figure}

Fig~\ref{labelnoise} summarizes the performance of each deep network trained using BUNDL and baseline methods across random, over- and under-segmented seizure labels and at -1.6-0.92 dB SNR. 
As seen, BUNDL consistently achieves the best performance across all noise mitigation strategies, as reflected in the window-level metrics of AUROC and AUPRC. 
For randomly noisy labels (rand), characterized by a symmetric mislabeling between the seizure and non-seizure classes, we observe improved overall performance across all metrics. The application of BUNDL leads to a significantly reduced false positive rate across all three models with marginal improvements in  sensitivity and latency. The NAL baseline performs comparably to BUNDL, while the SelfAdapt baseline and CEL, which assume no label noise, exhibit lower performance. Further, we discuss the effect of EEG SNR level and symmetric noisy labels (rand) on performance in  \nameref{S1_Appendix:snr} and \nameref{S1_snr}. Yet again, BUNDL showed consistent performance at all SNR levels compared to baseline methods indicating that baseline strategies are  likely latching on to noise to make their predictions.

%over-segmented seizure
In cases  with over-sampled seizures both at rates of 10\% (over 0.1) and 30\% (over 0.3), %The cases with 50\% over/under-sampling are provided in the appendix. 
the key challenge for the deep networks is to avoid learning from incorrectly labeled seizure classes, which in turn would lead to false-positive detections. As shown in Fig~\ref{labelnoise}, BUNDL achieves significantly lower FPR across all models and levels of over-segmented label noise while maintaining comparable sensitivity and latency. The NAL baseline performs second best overall, but its FPR improvement is inconsistent; for instance, with DeepSOZ at over 0.3, the FPR actually increases compared to CEL. SelfAdapt improves FPR and latency  at the cost of reduced sensitivity, indicating that it primarily lowers overall predicted probabilities without  addressing label noise.

%under-segmented
Finally, we also report the cases in which the seizure class is under-sampled at rates of 10\% (under 0.1) and 30\% (under 0.3). %Once again, the cases with 50\% under-sampling is provided in the appendix. 
Under-segmentation reflects a scenario where clinicians may not fully mark the seizure duration. Hence, the main challenge for the deep networks is to learn seizure characteristics with fewer samples 
BUNDL effectively maintains sensitivity even as training label noise increases from 10\% to 30\%, showing a significant improvement compared to baseline methods and CEL while maintaining comparable latency and FPR. In contrast, both SelfAdapt and CEL show a substantial drop in sensitivity as label noise levels increase, where the improvement seen with NAL over CEL is not significant.

\subsubsection*{{Ablation experiments}}

\begin{figure}[!h]
        \centering
        \includegraphics[width=0.85\linewidth]{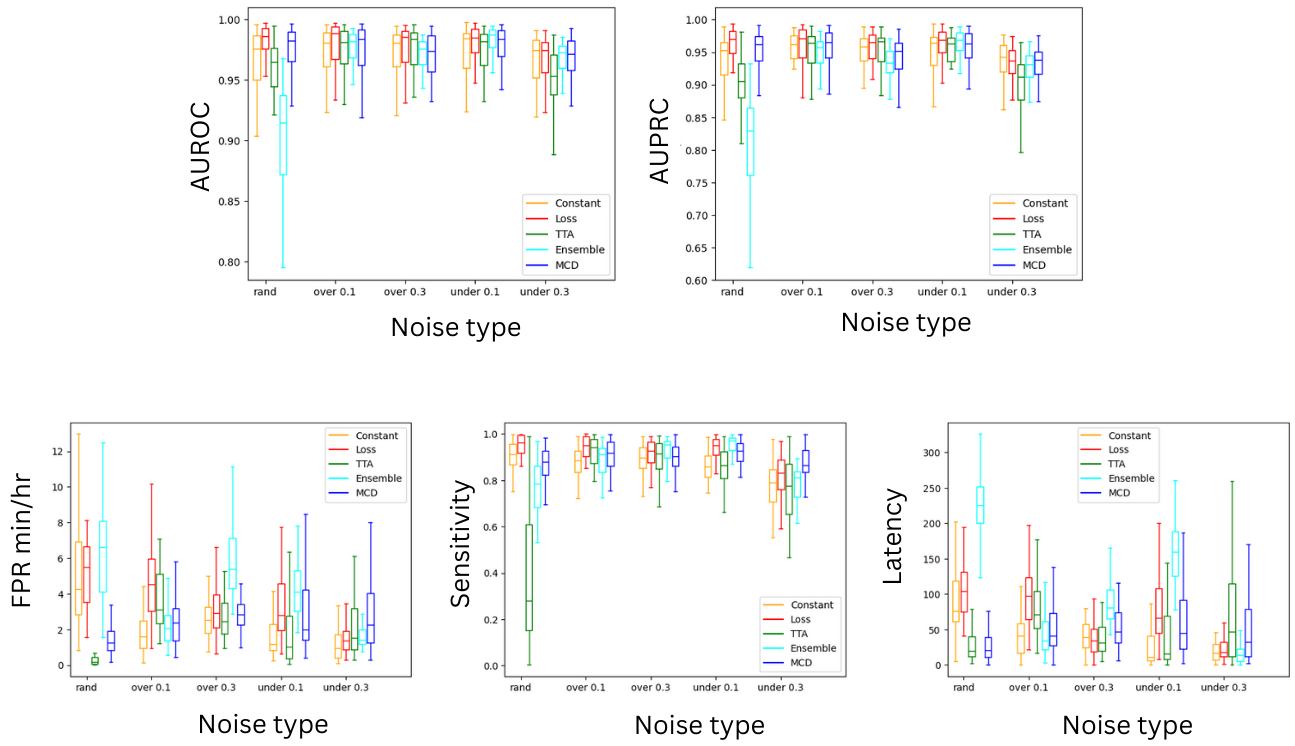}
        \caption{{\textbf{Seizure detection performance on simulated data for different uncertainty quantification methods using the DeepSOZ architecture.} Multiple metrics (AUROC, AUPRC, FPR min/hour, sensitivity, and latency in seconds) are shown. Three types of label noise settings are considered: rand - randomized symmetric, over 0.3 - 30\% over-segmentation, and under 0.3 - 30\% over-segmentation of seizures.}}
        \label{fig:ablations}
    \end{figure}

{Fig~\ref{fig:ablations} examines different uncertainty estimation methods under three types of label corruption (random symmetric noise, 10\% and 30\% over-segmentation, and 10\% and 30\% under-segmentation) using DeepSOZ. We replace the MCD-based entropy~$z_{xy_n}$ in the loss function with: (a) ensemble-based average entropy from outputs of five independently initialized models~\cite{borovac2022calibration} (random seeds of 42, 33, 0, 20, 25), (b) test-time augmentation (TTA) to generate multiple perturbed versions of each input for computing average entropy of output ~\cite{10340330,smith2025uncertainty}. Multiple input samples are generated by by adding Gaussian noise (mean 0, standard deviation 1) that is rescaled by 0.1, randomly flipping time windows, and rescaling in range (0.9, 1.1) to each of them before passing to  the model for computing average entropy of output probabilities, (c) batch-normalized cross-entropy loss, and (d) a constant value of 0.9.} 

{The results paint a nuanced comparison between the ablation methods. TTA and Ensemble based methods show significantly lower AUROC and AUPROC with randomized noisy label case accompanied by poor trade-off between FPR and sensitivity. The ensemble method is also  memory-intensive and show higher false positive rates. TTA also achieves lower sensitivity and considerably worse metrics in the random noise case. Loss-based and constant values achieve reasonable performances in the under-segmented seizure cases have higher false positive rates and latency in the random noise setting and over-segmented seizure cases. Moreover, unlike MCD, Loss and Constant values do not show any difference in their average values when labels are correct or incorrect as shown in Table \ref{tab:uqrange}. Thus, we conclude that MCD-based uncertainty offers the most effective and reliable way to identify label mismatch for refinement in seizure detection.}

\subsection*{Real world data}

Unlike the simulated data, we do not have access to the ground-truth ``clean" seizure labels in  real-world datasets. Hence, the model performances are evaluated against the clinician-provided (and possibly noisy) labels. In addition, we present the results from training model with the knowledge of over-segmented seizures. This hypothesis is validated by several articles discussing the pitfalls of inter-rater variability and over-segmented seizures~\cite{amin2019role,halford2015inter}. Hence, the main challenge is to reduce the false positive predictions that the deep networks would learn from the increased (and possibly faulty) representation of seizure class labels in the annotations.

%\medskip \noindent
{Tables~\ref{tab:tuh} and \ref{tab:chb}  summarizes the AUROC score and seizure-level performances of BUNDL and the baseline label noise methods across three deep networks on TUH and CHB-MIT datasets, respectively. Results on the TUH dataset using DeepSOZ model were obtained by re-implementing our prior work in~\cite{shama2024uncertainty}; the additional deep networks and baseline methods are unique to this study. As seen, BUNDL consistently achieves the lowest false positive rate (FPR) with the top performance using DeepSOZ on the TUH dataset (2.3 min/hour) and using the CNN on CHB-MIT (0.8 min/hour). The improvement in FPR with BUNDL over NAL and CEL is statistically significant using the best performing model on each dataset. BUNDL also achieves statistically significant improvement in latency over the baseline methods in most cases across both datasets. There is no statistically significant improvement in AUROC scores except in two model configurations.}

    \begin{table*}[!ht]
    \begin{adjustwidth}{-0.95in}{0in} % Comment out/remove adjustwidth environment if table fits in text column.
        \centering
        \caption{\textbf{Comparison of BUNDL and baseline methods on the TUH dataset using three deep networks}. }% For HViT, p-values are computed against BUNDL for all three networks.}}

        \label{tab:tuh}
         \begin{tabular}{|c|c|c|c|c|c|}
    \hline
    \multirow{2}{*}{Deep network} & \multirow{2}{*}{Method}  &\multicolumn{4}{c|}{\textbf{TUH} } \\ \cline{3-6} 
      &     & {AUROC} & FPR  & Sensitivity & Latency  \\ \hline

    \hline
    \hline
      {\multirow{4}{*}{DeepSOZ\cite{m2023deepsoz}}} &  BUNDL\textdagger & {\textbf{0.917 $\pm$ 0.031}} &\textbf{2.30$\pm$1.10} & 0.74$\pm$0.08 & \textbf{7.6$\pm$10.9}  \\ %\cline{2-8} 
            &   Selfadapt\cite{huang2020self}\textdagger & {0.908 $\pm$ 0.03}  &2.50$\pm$1.23& 0.77$\pm$0.07 & 17.7$\pm$14.8 *   \\ %\cline{2-8} 
            &   NAL\cite{goldberger2016training}\textdagger &  {0.912 $\pm$ 0.034}   &4.24$\pm$1.96 *&  \textbf{0.83$\pm$ 0.06} * & 22.3$\pm$ 15.1 *  \\ %\cline{2-8} 
            &   CEL\textdagger & {0.916 $\pm$ 0.028}  & 3.6$\pm$1.77 *  & 0.79$\pm$0.06 * & 16.9$\pm$16.2 * \\ 
                    
    \hline
    \hline
    
     {\multirow{4}{*}{TGCN\cite{covert2019temporal}}} &  BUNDL &  {0.891 $\pm$ 0.036} &\textbf{2.6$\pm$2.5 } &  0.75$\pm$0.09 &\textbf{30.1$\pm$29.7}   \\ %\cline{2-8} 
            &   Selfadapt\cite{huang2020self} & {0.89 $\pm$ 0.033} &3.0$\pm$1.2   &  0.81$\pm$0.08 *    &   41.2$\pm$17.3 *  \\ %\cline{2-8} 
            &   NAL\cite{goldberger2016training}  & {\textbf{0.899 $\pm$ 0.031}} & 2.8$\pm$1.5  &  \textbf{ 0.87$\pm$0.07} * &  50.4$\pm$28.5 *\\ %\cline{2-8} 
            &   CEL & {0.898 $\pm$ 0.032 }& 3.3$\pm$1.6  & 0.84$\pm$0.06 *&  41.3$\pm$19.5 *   \\ 
                    
    \hline
    \hline
    
      {\multirow{4}{*}{{CNN\cite{craley2021automated}}} } &  BUNDL &{0.882 $\pm$ 0.057}&\textbf{ 2.7$\pm$3.5}  & 0.54$\pm$0.23 &\textbf{11.2$\pm$28.9} \\ %\cline{2-8} 
            &   Selfadapt\cite{huang2020self} & {0.891 $\pm$ 0.045} &3.4$\pm$1.9  & 0.66$\pm$0.09 *  &   15.7$\pm$16.1\\ %\cline{2-8} 
            &   NAL\cite{goldberger2016training} & {0.879 $\pm$ 0.065} &5.6$\pm$3.3 *    & \textbf{0.85$\pm$0.11} *  &   43.7$\pm$33.0 *\\ %\cline{2-8} 
            &   CEL & {\textbf{0.896 $\pm$ 0.034}}  &3.0$\pm$2.2 & 0.66$\pm$0.13 * & 15.5$\pm$20.1   \\ 
     
    \hline
    {HViT} \cite{deng2023eeg} & {DUL}\cite{deng2023eeg} &  {0.912 $\pm$ 0.031} &  {2.30 $\pm$ 0.7} &  {\textbf{0.912 $\pm$ 0.077}} * &{58.4 $\pm$ 16.8 *}\\ \hline
    % \begin{flushleft} Table notes \end{flushleft}
    \end{tabular}
        \begin{flushleft}
        Results show mean and standard deviation over repeated 10-fold cross-validation. Best per network is in bold; best overall is underlined.
        \textit{*Indicates significantly different (p$<$0.05) compared to BUNDL.} \textdagger \textit{Reimplemented from \cite{shama2024uncertainty}}
    \end{flushleft}
    \end{adjustwidth}
    \end{table*}

%\\ \vspace{0.5in}

    \begin{table*}[!ht]
    \begin{adjustwidth}{-0.95in}{0in} % Comment out/remove adjustwidth environment if table fits in text column.
        \centering
        \caption{\textbf{Comparison of BUNDL and baseline methods on the CHB-MIT dataset using three deep networks.} } % For HViT, p-values are computed against BUNDL for all three networks.}}

        \label{tab:chb}
         \begin{tabular}{|c|c|c|c|c|c|}
    \hline
    \multirow{2}{*}{Deep network} & \multirow{2}{*}{Method}  & \multicolumn{4}{c|}{\textbf{CHB-MIT}}  \\ \cline{3-6} 
      &     & {AUROC} & FPR  & Sensitivity & Latency  \\ \hline
               
    \hline
    \hline
      {\multirow{4}{*}{DeepSOZ\cite{m2023deepsoz}}} &  BUNDL &  {\textbf{0.865 $\pm$ 0.076}} &\textbf{1.7$\pm$2.9 } & 0.58$\pm$0.28 &   \textbf{21.54$\pm$56.8} \\ %\cline{2-8} 
            &   Selfadapt\cite{huang2020self} &  {0.781 $\pm$ 0.117} *  & 2.2$\pm$5.2  & 0.41$\pm$0.30 * & 28.3$\pm$47.1  \\ %\cline{2-8} 
            &   NAL\cite{goldberger2016training} & {0.869 $\pm$ 0.09}  & 3.4$\pm$4.6 * & 0.59$\pm$0.28  &  46.8$\pm$94.6   \\ %\cline{2-8} 
            &   CEL &{0.864 $\pm$ 0.089} & 4.4$\pm$5.7 * & \textbf{0.63$\pm$0.23} & 79.7$\pm$167.1 * \\ 
                    
    \hline
    \hline
    
     {\multirow{4}{*}{TGCN\cite{covert2019temporal}}} &  BUNDL & {0.859 $\pm$ 0.109} & \textbf{ 0.8$\pm$1.8}  &  0.71$\pm$0.29 &  \textbf{14.7$\pm$11.6}  \\ %\cline{2-8} 
            &   Selfadapt\cite{huang2020self} &  {0.860 $\pm$ 0.132} &2.6$\pm$8.1   &    0.75$\pm$0.26   &  51.9$\pm$ 264.2 \\ %\cline{2-8} 
            &   NAL\cite{goldberger2016training}  &  { 0.804 $\pm$ 0.121} *   &0.84$\pm$0.28  &  0.77$\pm$0.24   &       18.1$\pm$12.2 *        \\ %\cline{2-8} 
            &   CEL &  {\textbf{0.884 $\pm$ 0.107}}  &3.2$\pm$1.0 *   &  \textbf{0.83$\pm$0.23 } * & 33.7$\pm$19.8 *  \\ 
                    
    \hline
    \hline
    
      {\multirow{4}{*}{{CNN\cite{craley2021automated}}} } &  BUNDL & {0.875 $\pm$ 0.128} &\textbf{0.8$\pm$0.7} &  0.64$\pm$0.33 & \textbf{ 10.2$\pm$11.6}   \\ %\cline{2-8} 
            &   Selfadapt\cite{huang2020self} & {\textbf{0.898 $\pm$ 0.113}} &1.0$\pm$0.7   &  0.67$\pm$0.33   &  11.0$\pm$7.0 \\ %\cline{2-8} 
            &   NAL\cite{goldberger2016training} &  {0.885 $\pm$ 0.129}  & 2.6$\pm$2.3 * &   0.72$\pm$0.31 & 21.8$\pm$15.5 * \\ %\cline{2-8} 
            &   CEL &  {0.887 $\pm$ 0.121}     &   2.7$\pm$1.2 *   & \textbf{ 0.76$\pm$0.28 }   &    28.7$\pm$82.3     \\ 
     
    \hline
    {HViT} \cite{deng2023eeg} & {DUL}\cite{deng2023eeg} & {0.831 $\pm$ 0.111} & {2.24 $\pm$ 3.34 *} & {0.590 $\pm$ 0.223}  & {62.2 $\pm$ 22.9 *}\\ \hline

    % \begin{flushleft} Table notes \end{flushleft}
    \end{tabular}
    \begin{flushleft}
        Results show mean and standard deviation over repeated 10-fold cross-validation. Best per network is in bold; best overall is underlined.
        \textit{*Indicates significantly different (p$<$0.05) compared to BUNDL.}
    \end{flushleft}
    \end{adjustwidth}
    \end{table*}
    
{The HViT-DUL baseline shows the highest sensitivity and similar FPR as that of BUNDL when applied to the TUH dataset. However, on the CHB-MIT dataset, it shows similar sensitivity with higher FPR than BUNDL. In both cases, HViT has high latency ($\sim1$ min), which is clinically undesirable. SelfAdapt has similar FPR and sensitivity scores with higher latency scores in many cases. On the other hand, NAL and CEL also achieve the moderately high sensitivity on CHB-MIT and significantly higher sensitivity when applied to the TUH dataset; both do so at the cost of significantly higher FPR and latency. In contrast, BUNDL’s improved FPR and latency with lower sensitivity may be expected given the possibly over-segmented clinicians annotations, thus underscoring the need for cleaner seizure demarcations in future work.}

\subsubsection*{Generalizability test}

{Unlike CHB-MIT, which uses bipolar referencing, the TUH and Siena datasets are collected in different hospitals but share similar recording protocols and average-referenced EEG, making cross-site testing appropriate. As shown in Table~\ref{tab:siena}, BUNDL achieves the lowest false positive rate (FPR) across all noisy label methods when evaluated against the seizure annotations. Again, we note that the Siena dataset is clinician-annotated, and some over-segmentation of seizure intervals is expected. The significant reduction in FPR suggests an improvement in handling potential label noise. We also observe a significant reduction in sensitivity with BUNDL relative to the baseline CEL, which may reflect both the over-segmentation in the Siena labels and the generalization challenges of deep networks across independent sites. SelfAdapt shows performance comparable to BUNDL, with variations depending on the underlying model:  improvement for CNNs and lower performance with DeepSOZ and TGCN. In contrast, NAL achieves high AUROC and sensitivity at the cost of an FPR exceeding 4 min/hour, indicating that it may overfit to potentially over-segmented seizure labels.  Similary, HViT-DUL \cite{deng2023eeg} shows the best sensitivity and latency however with extremely high FPR or 30.2 min/hour indicating that not accounting for noisy labels during training can lead to overfitting to possibly over-segmented seizures. The baseline CEL with no label noise correction maintains a reasonable balance between sensitivity and FPR, though its performance is reduced from the TUH dataset where models were trained. Nonetheless, both BUNDL and SelfAdapt show promise in cross-site generalization, which can be explored in future domain adaptation work if the drop in sensitivity can be eliminated using test-time adaptation.}

        \begin{table*}[!ht]
        \begin{adjustwidth}{-1.in}{0in}
     \centering
     %\color{black}
     
        \caption{{\textbf{Comparison of BUNDL and baseline methods using three deep networks on the unseen Siena dataset.}}} %p-value for HViT is computed against BUNDL with all three deep networks.}}
        \label{tab:siena}
        \begin{tabular}{|ll|l|l|l|l|}
        \hline
        \multicolumn{2}{|l|}{}                                      & AUROC                                   & FPR                                              & Sens                                    & Latency      \\ \hline
        \multicolumn{1}{|l|}{}                          & BUNDL     & 0.887$\pm$ 0.010 &                  \textbf{1.465$\pm$ 0.203}                         &0.446$\pm$ 0.042                          & \textbf{192.4$\pm$ 18.7} \\
        \multicolumn{1}{|l|}{}                          & SelfAdapt & 0.869 $\pm$ 0.009    *                   & 1.743 $\pm$ 0.293 *                         & 0.458 $\pm$ 0.049                       & 195.4 $\pm$ 8.3     \\
        \multicolumn{1}{|l|}{}                          & NAL       & \textbf{0.901 $\pm$ 0.015}*              & 4.956 $\pm$ 1.785  *                         & \textbf{0.696 $\pm$ 0.052} *             & 202.8 $\pm$ 21.9 *\\ 
        \multicolumn{1}{|l|}{\multirow{-4}{*}{DeepSOZ}} & CEL       & 0.898 $\pm$ 0.011 *                     & 2.331 $\pm$ 1.167 *                        & 0.535 $\pm$ 0.062 *                  & 198.2 $\pm$ 12.5   \\ \hline
        \multicolumn{1}{|l|}{}                          & BUNDL     & 0.830 $\pm$ 0.011                       & \textbf{4.503 $\pm$ 0.501}                       & 0.510 $\pm$ 0.021                       & 111.5 $\pm$ 6.7    \\
        \multicolumn{1}{|l|}{}                          & SelfAdapt & 0.812 $\pm$ 0.066                       & { 5.085 $\pm$ 0.296}*                           & 0.477 $\pm$ 0.024 *                      & 97.3$\pm$ 4.3     * \\
        \multicolumn{1}{|l|}{}                          & NAL       & \textbf{0.842 $\pm$ 0.011} *             & 5.670 $\pm$ 0.040  *                   & 0.588 $\pm$ 0.039 *                      & \textbf{73.0 $\pm$ 9.51}*   \\
        \multicolumn{1}{|l|}{\multirow{-4}{*}{TGCN}}    & CEL       & 0.830 $\pm$ 0.012                       & 7.560 $\pm$ 0.640  *             & \textbf{0.619 $\pm$ 0.027} *                       & 75.2 $\pm$ 9.03 * \\ \hline
        \multicolumn{1}{|l|}{}                          & BUNDL     & 0.883 $\pm$ 0.06                        & {\ul{\textbf{0.314 $\pm$ 0.215}}}                 & 0.377 $\pm$ 0.157                       & 234.1 $\pm$ 46.7   \\
        \multicolumn{1}{|l|}{}                          & SelfAdapt & {\ul{\textbf{0.947 $\pm$ 0.01}}} *        & 1.64 $\pm$ 1.114  *                               & 0.742 $\pm$ 0.04 *                       & 226.8 $\pm$ 17.3  \\
        \multicolumn{1}{|l|}{}                          & NAL       & 0.857 $\pm$ 0.215                       & 10.49 $\pm$ 4.79 *                          & \textbf{0.749 $\pm$ 0.088}  *          & \textbf{92.6 $\pm$ 52.9}* \\
        \multicolumn{1}{|l|}{\multirow{-4}{*}{CNN}}     & CEL       & 0.896 $\pm$ 0.033                       & 1.86 $\pm$ 1.08  *                    & 0.615 $\pm$ 0.091   *                   & 238.6 $\pm$ 42.6     \\ \hline
        \multicolumn{1}{|l|}{HViT}                      & DUL       & 0.850 $\pm$ 0.023 *                      & 30.2 $\pm$ 9.55 *                    & {\ul{\textbf{0.896 $\pm$ 0.034}}} *                & {\ul{\textbf{14.7 $\pm$ 12.2}}} *   \\ \hline
        \end{tabular}
    \begin{flushleft}
        Mean and standard deviation is across repeated cross validation folds are reported. Highest per deep network is indicated in bold and highest across the table is indicated in underline. \textit{*Indicates significantly different (p$<$0.05) compared to BUNDL.}
    \end{flushleft}
        \end{adjustwidth}
    \end{table*}
\subsubsection*{Uncertainty estimation and label transition in BUNDL}

\begin{figure}[!h]
    %\centering
    \includegraphics[width=0.96\linewidth]{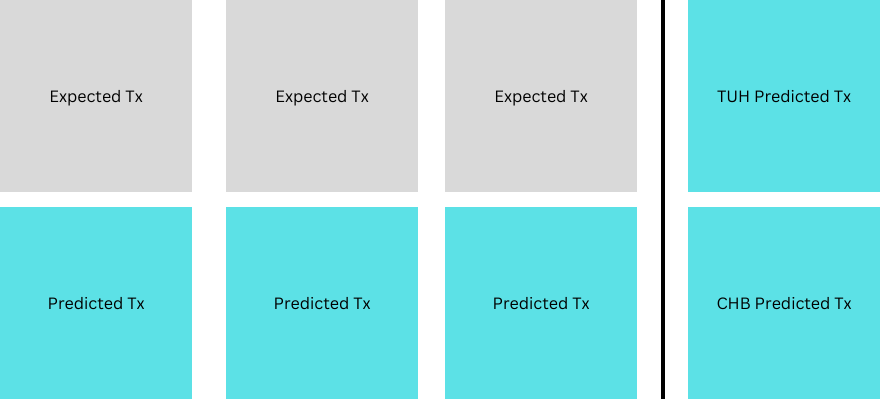}
    \caption{\textbf{Computed uncertainty levels and  transition matrices by DeepSOZ - BUNDL in  simulated and real datasets.}}
    Each sub-figure includes \textbf{Left:} Boxplots that show computed uncertainty within 1 minute of inter-ictal, preictal, and ictal periods, with mean and variance across cross-validation folds; and \textbf{Right:}. The corresponding preictal transition matrices given by Eq.~(\ref{eqn:cleanpost}). Higher uncertainty leads to greater annotation ambiguity, reflected by more white (non-diagonal) values in the transition matrix. %With no uncertainty, the matrix would be purely diagonal.  }
    \label{trans}
\end{figure}

We evaluated data-dependent label transitions predicted by DeepSOZ trained using BUNDL (DeepSOZ-BUNDL) to analyze the distributions learned by the model in different scenarios. For each subject, we randomly selected one-minute segments from interictal ($>$2 minutes before onset or after offset), preictal (around onset), and ictal (during seizure) periods. At these time points, we computed the BUNDL uncertainty metric $z_{xy_n}$ (Eq.~\ref{eqn:unc}), averaged across time and subjects per cross-validation fold. As shown in Fig~\ref{trans}, the uncertainty was significantly higher in preictal and ictal periods across all folds. Among the datasets, CHB-MIT exhibited lower overall uncertainty, while TUH was substantially higher. These findings align with our simulation setup, where noise was added near onset, and with real-world patterns of increased uncertainty around changepoints (preictal) and complex ictal patterns.

As further validation, we assess the conditional distribution$ p(y_c/ y_n, x)$ given by Eq.~(\ref{eqn:cleanpost}) to examine the transition from the noisy label $y_n$ to the true label $y_c$ in the BUNDL-DeepSOZ model. This analysis is averaged over preictal time points (one minute around onet) in \(y_n\). As shown in Fig~\ref{trans} the probability of \(y_c=1\) given \(y_n=0\) for under-segmented seizures in the simulated dataset increases from 0.28 at 10\% label noise to 0.34 as the label noise level rises to 30\%. This observation aligns with the simulation setup, i.e., the seizure duration shortens with increasing label noise, meaning more non-seizure points in \(y_n\) overlap with the true seizure \(y_c\). Consequently, as label noise increases, \( p(y_c=1 \mid y_n=0, x) \) also increases. Moreover, \( p(y_c=1 / y_n=1, x) \) remains high because the under-segmented seizure in \(y_n\) coincides with the true seizure \(y_c\). In contrast, for over-segmented training labels, the non-seizure points in \(y_n\) have low uncertainty by design and align with true non-seizure points in \(y_c\), keeping \( p(y_c=0 \mid y_n=0, x) \) close to 1. However, \( p(y_c=1 / y_n=1, x) \) decreases as label noise increases, since a larger portion of the noisy seizure in \(y_n\) no longer corresponds to the true seizure in \(y_c\). BUNDL effectively captures these label mismatches introduced during simulation. In the TUH and CHB-MIT datasets, where only noisy labels are available, the predicted distributions exhibit trends consistent with over-segmented seizures. For TUH, \( p(y_c=1 / y_n=1, x) \) is around 0.62, indicating that approximately 62\% of clinician annotations align with true seizures. For CHB-MIT, this metric is around 0.75, which corresponds to lower uncertainty levels and suggests better annotation quality in the dataset.

\subsubsection*{SOZ localization in TUH}

The DeepSOZ architecture was developed for the dual tasks of seizure detection and onset zone localization~\cite{m2023deepsoz}. To assess whether accounting for noisy labels enhances the latter task, we trained the localization branch of DeepSOZ after the detection had been trained using BUNDL. This model is compared to naive training of DeepSOZ with CEL, i.e., not accounting for noisy detection labels. At the seizure level, BUNDL achieved an accuracy of 0.633$\pm$0.149, compared to 0.601$\pm$0.176 with CEL. At the patient level, where localization predictions are averaged across all recordings for a comprehensive output, BUNDL improved the accuracy from 0.591$\pm$0.144 (CEL) to 0.620$\pm$0.111. We hypothesize that the reduction in false positives in seizure detection also reduces false onset zone predictions as seen in two example patients in as discussed in \cite{shama2024uncertainty}. This is likely due to the connection between seizure detection and onset zone localization in the attention-based pooling layer of DeepSOZ. Therefore, incorporating BUNDL enhances the overall effectiveness of a dual-task epilepsy monitoring system.

\section*{Discussion}
%summary
We have introduced BUNDL as an effective solution for training deep networks with noisy labels, specifically  applied to epileptic seizure detection. BUNDL leverages a Monte Carlo (MC) dropout procedure to compute uncertainty and adaptively adjust the loss function during training with noisy labels. Analysis on simulated and real-world EEG data (TUH and CHB-MIT) demonstrates BUNDL's robustness in identifying noisy samples and accurately learning seizure characteristics, with consistent performance improvements. Notably, using BUNDL to improve  detection also improves the downstream task of seizure onset localization in TUH. Overall, our findings highlight BUNDL's benefits in improving the reliability of automated epilepsy management.

\subsection*{Uncertainty quantification as a label noise measure}
% model agnosticity and reducing computational complexities. 
We believe the most significant contribution of this work lies in the efficiency and model-agnosticity enabled by its underlying Bayesian approach. By modeling uncertainty through a simple yet effective MC dropout procedure, BUNDL can implicitly identify unreliable annotations during training without requiring architectural changes or additional memory. This efficiency lies in stark contrast to baselines such as NAL and Self-Adapt, which introduce learnable parameters and/or must explicitly store past information. In addition, the MC dropout is built into the proposed loss function and does not rely on post hoc adjustments. While MC dropout is generally slower than a gradient update, the availability of GPUs during training allows for parallel or batch processing of each forward pass, effectively mitigating the time cost. Moreover, MC dropout is only required during training; during testing, BUNDL operates without any additional steps, maintaining the same efficiency as the traditional cross-entropy loss (CEL) that assumes no noise. BUNDL also delivers consistently strong performance across three deep networks due to its model-agnosticity in multiple datasets, with steady improvement during training. Its uncertainty estimates help justify deviations from noisy labels, effectively enabling a built-in quality check.

%para on improvement over baselines
In contrast, the popular NAL approach captures model uncertainty by introducing additional parameters that scale quadratically with the number of classes, requires a modified loss function, and involves a three-stage training process—making it both cumbersome and computationally inefficient. Despite these complexities, NAL generally performs worse than BUNDL. SelfAdapt avoids the parameter overhead but  requires extra memory to store past model outputs. Our experiments show that it struggles with noisy labels in complex datasets like EEG and tends to reduce overall prediction confidence, as it lacks an internal mechanism for estimating uncertainty. In comparison, BUNDL provides a more principled and efficient solution to the pervasive problem of label noise, particularly in visually complex, manually annotated domains like EEG.

{The HViT-DUL baseline models noise in the EEG data through its parameters, which is distinct from BUNDL, which captures \textit{label uncertainty}. Nonetheless, HViT-DUL achieved commendable performance on the TUH dataset with similar AUROC and FPR profiles as BUNDL-DeepSOZ and a significantly higher sensitivity. However, these performance gains are offset by significantly higher latency. These trends are not consistent across datasets. On CHB-MIT, HViT-DUL has similar sensitivity and higher FPR, as compared to BUNDL. The cross-site generalization performance on Siena is also worse, as the FPR of HViT-DUL increased to roughly 30 min/hour, which is clinically undesirable. Thus, we conclude that despite its advantages in uncertainty modeling for EEG data, HViT-DUL could benefit from directly handling label noise.}
% para on how EEG dynamics are captured and BUNDL understands seizure propogation. "data dependent transition" application specific

\subsection*{Robustness and generalizability}

In addition to the technological innovation of BUNDL, we introduce a simulated EEG dataset designed to benchmark noisy label algorithms under five structured noise scenarios, including both symmetric and asymmetric cases. This dataset enables fair comparisons across methods, and the provided code allows easy extension to other scenarios. In both simulated and real-world datasets, we demonstrate how BUNDL effectively incorporates domain knowledge about annotation uncertainties during training. By leveraging existing seizure detection networks, BUNDL captures EEG dynamics while estimating the label uncertainty through the novel Bayesian framework. As shown in our results, the estimated uncertainty estimation in BUNDL aligns well with clinically ambiguous periods, such as preictal and early ictal phases, where seizure propagation is often unclear. Importantly, BUNDL uses this uncertainty to refine seizure detection outputs, leading to improved performance in downstream task of seizure onset zone localization. 

{Table~\ref{tab:tflops} compares the computational cost of BUNDL with baseline algorithms across the three deep networks. TFLOPs depend on the size of the base deep network and the input dimensionality, particularly for convolutional and transformer-based architectures~\cite{sovrasov2018ptflops,vaswani2017attention}. They also scale linearly with dataset size, number of training epochs, and the number of training stages (e.g., pretraining and finetuning). Across all three deep networks, the CEL method without any noisy-label correction has the lowest computational cost, as it involves only a single training stage, requiring 27 TFLOPs, 170 TFLOPs, and 5 TFLOPs for DeepSoz, TGCN, and CNN-BLSTM, respectively, on the TUH dataset. HViT, another baseline with no label noise correction, also has a lower computational cost. Since all label noise learning strategies begin with pretraining, they add to the computation cost of CEL training. BUNDL and SelfAdapt require an additional cost of 290, 1879, and 50 TFLOPs on the three models, respectively, due to averaging over multiple model predictions. While BUNDL uses batch processing to run multiple forward passes using dropout, SelfAdapt stores multiple models from previous training iterations to speed up computation, thus trading off time for increased memory usage. While NAL adds a smaller number of TFLOPs during test-time inference, it requires 2 additional training loops to be done sequentially after pretraining, thus adding to the total training time. Similar trends are observed on the CHB-MIT dataset, which has fewer EEG recordings than TUH but with longer inputs of 60 minutes length.}

\begin{table*}[!ht]
%\color{black}
    \begin{adjustwidth}{-1.in}{0in} % Comment out/remove adjustwidth environment if table fits in text column.
    \centering
    \caption{{\textbf{ Comparison of the computational cost of BUNDL with baseline algorithms across the three deep networks.} }}
    \label{tab:tflops}
    \begin{tabular}{|ll|l|l|l|l|}
    \hline
    \multicolumn{2}{|l|}{}                                 & \begin{tabular}[c]{@{}l@{}}DeepSOZ \\ Params = 510K\end{tabular} & \begin{tabular}[c]{@{}l@{}}TGCN \\ Params = 1.16M\end{tabular} & \begin{tabular}[c]{@{}l@{}}CNN\\ Params = 30K\end{tabular} & \begin{tabular}[c]{@{}l@{}}HViT-DUL\\ Params = 246K\end{tabular} \\ \hline
    \multicolumn{1}{|l|}{\multirow{4}{*}{TUH}} & BUNDL     & 1$\times$27 + 1$\times$290                                                     & 1$\times$170 + 1$\times$1879                                                 & 1$\times$5 + 1$\times$50                                                 & –                                                                \\ \cline{2-6} 
    \multicolumn{1}{|l|}{}                     & SelfAdapt & 1$\times$27 + 1$\times$290                                                     & 1$\times$170 + 1$\times$1879                                                 & 1$\times$5 + 1$\times$50                                                 & –                                                                \\ \cline{2-6} 
    \multicolumn{1}{|l|}{}                     & NAL       & 1$\times$27 + 2$\times$28                                                      & 1$\times$170 + 2$\times$171                                                  & 1$\times$5 + 1$\times$6                                                  & –                                                                \\ \cline{2-6} 
    \multicolumn{1}{|l|}{}                     & CEL/DUL*  & 1$\times$27                                                             & 1$\times$170                                                          & 1$\times$5                                                        & 1$\times$6                                                        \\ \hline
    \multicolumn{1}{|l|}{\multirow{4}{*}{CHB-MIT}} & BUNDL     & 1$\times$58 + 1$\times$644                                                     & 1$\times$378 + 1$\times$4158                                                 & 1$\times$10 +1$\times$110                                                &                                                                  \\ \cline{2-6} 
    \multicolumn{1}{|l|}{}                     & SelfAdapt & 1$\times$58 + 1$\times$644                                                     & 1$\times$378 + 1$\times$4158                                                 & 1$\times$10 + 1$\times$110                                               &                                                                  \\ \cline{2-6} 
    \multicolumn{1}{|l|}{}                     & NAL       & 1$\times$58 + 2$\times$59                                                      & 1$\times$378 + 2$\times$379                                                  & 1$\times$10 + 2$\times$11                                                &                                                                  \\ \cline{2-6} 
    \multicolumn{1}{|l|}{}                     & CEL/DUL * & 1$\times$58                                                             & 1$\times$378                                                          & 1$\times$10                                                       & 1$\times$11                                                       \\ \hline
    \end{tabular}
    \begin{flushleft}
        The values are presented as follows: (number of pretaining stages $\times$ training TFLOPS + number of finetuning stages $\times$ training TFLOPS). *\textit{Not a noisy label learning method.}
    \end{flushleft}
    \end{adjustwidth}
    \end{table*}

{In summary, BUNDL presents a novel training algorithm and a distributional loss to deliver robust and generalizable end-to-end seizure detection models. BUNDL comes with an inference time of approximately 0.11 seconds with a GPU A100 and 3 seconds with a CPU for detection results enabling real-time clinical deployment. Furthermore, BUNDL can be adapted to other imaging modalities due to its model-agnostic nature. One of BUNDL’s key strengths is its ability to incorporate annotations from multiple raters or clinicians by combining their scores within the noisy label distribution parameter. This capability makes BUNDL particularly suitable for medical applications, where inter-rater variability is common and accurate label estimation is crucial.}

\subsection*{Limitations and future work}

{The main goal of  BUNDL is to estimate and correct for unknown label noise when training deep learning models. This procedure, in turn, should improve performance, as compared to the standard CEL. We emphasize that BUNDL is not being proposed as a novel seizure detection model but rather as a refinement algorithm. This fact also highlights a limitation of the current study, and of most label noise mitigation strategies that rely on pretrained model components. Pretraining is a common and often necessary step in noisy-label learning to first capture the underlying task and obtain well-calibrated label transition probabilities \cite{northcutt2021confident,huang2020self,goldberger2016training}, including the uncertainty estimates used in our method. For instance, the baseline methods of noisy label learning like Selfadapt\cite{huang2020self} includes pretraining once before refinement, and NAL\cite{goldberger2016training} includes two steps of pretraining for base model and noise layers respectively. A promising future direction for this area of research is to develop prior distributions that more correctly reflect the underlying data uncertainties, thereby providing a set of regularization constraints to the loss function during training and eliminating the need for pretraining. In addition, integrating self-supervised training methods along with BUNDL may allow the learning of robust features  by correctly penalizing the training procedure and helping to avoid any performance loss due to incorrect labels.}  
    
{Moreover, BUNDL is a training algorithm that can be applied to any existing model by leveraging MCD-based uncertainty quantification without altering the underlying network, thereby maintaining model agnosticity. However, our use of MCD prevents exploration of more robust uncertainty-estimation techniques proposed in the Bayesian neural network literature \cite{deng2023eeg}. MCD-based uncertainty estimation also requires multiple forward passes, which can be computationally expensive (Table~\ref{tab:tflops}). On the other hand, modeling epistemic and aleatoric uncertainty through added parameters for learned data priors introduces architectural changes and requires retraining. While we have intentionally chosen not to alter the structure of existing deep learning models to enable seamless integration of BUNDL with established work, future studies should focus on coupling the model structure and uncertainty quantification \cite{kendall2017uncertainties,li2022eegunc,deng2023eeg} when using BUNDL to adjust for noisy labels during training.}
    
{With respect to cross-site generalization, BUNDL achieves the lowest false-positive rate when evaluated on the unseen Siena dataset, indicating its effectiveness in mitigating the impact of noisy labels. However, this improvement was accompanied by a significant drop in sensitivity. Two factors may explain this outcome: class imbalance that can hinder training with BUNDL, and the overall limited capacity of the models, as all methods exhibited performance degradation on this out-of-distribution dataset. Since uncertainty quantification is not applied during testing, this decline reflects the challenges deep networks face when transferring to new clinical domains. Future work may explore self-supervised approaches combined with uncertainty-aware test-time adaptation to improve sensitivity while maintaining tolerance to potential label noise~\cite{li2023t,wang2025neurottt,mao2023online}.}  
    
{Finally, while our experiments show potential for real-time adaptation of BUNDL-trained models, validation in a clinical environment with real-time experiments is crucial. The fundamental challenge to real-world validation of noisy label methods is that we lack ground-truth seizure annotations. The performance metrics that we report for the real-world datasets were computed based on the clinician-provided seizure annotations, which are likely to be imprecise \cite{halford2013standardized,halford2015inter}. As noted in the literature~\cite{halford2013standardized,halford2015inter,jing2020interrater}, clinicans are likely to over-segment the seizures in an effort to avoid false negative detections. Thus, some drop in sensitivity in TUH and CHB-MIT datasets may be partially explained by the possibly imprecise labels and the class imbalance. Including annotations from multiple clinicians and also factoring in rater confidence during training can facilitate better models and evaluation strategies to combat label noise. The computational efficiency of BUNDL paves the way for future studies that explore real-time clinical adaptation to epilepsy monitoring. Future work could also go beyond epilepsy and apply BUNDL to other medical domains that rely on manual annotations, such as pathology and radiology, ultimately contributing to more reliable deep learning for healthcare.}

\section*{Conclusion}
We have introduced BUNDL, a Bayesian strategy to correct for noisy labels when training deep networks. By introducing MC dropout and a novel loss function, BUNDL  estimates the label uncertainty and adjusts the learning process, reducing the risk of the model learning from incorrect features. Our experiments across multiple datasets, %including both simulated and real EEG data, 
consistently demonstrated BUNDL's ability to improve key performance metrics such as false positive rates and latency in seizure detection. Unlike other approaches, BUNDL requires no additional parameter overhead, making it a resource-efficient solution with  strong performance. Moreover, BUNDL's versatile, model-agnostic framework offers a robust solution for handling noisy labels across domains, enhancing deep learning performance and reliability in critical medical applications.
%\newpage

\subsection*{{Data availability statement}}

{Our scripts to generate simulated data, training, and evaluation are available on \href{https://github.com/deeksha-ms/BUNDL}{Github.}. Real world datasets are publicly available in their original websites \cite{shah2018temple,guttag2010chb,detti2020siena}}.

\section*{Supporting information}

% Include only the SI item label in the paragraph heading. Use the \nameref{label} command to cite SI items in the text.

\paragraph*{S1 Appendix 1.} 
\label{S1_Appendix:snr}
{\textbf{Effect of EEG signal-to-noise ratio in the performance of BUNDL and other noisy label algorithms.}}
{We hypothesize that increased noise in EEG correlates with label ambiguities and further makes training harder as modeled in our graphical model in Fig~\ref{train}(left).
Thus, to evaluate the effect of signal-to-noise ratio (SNR) on model performance, we  introduced Gaussian noise at three different amplitude levels during the creation of our simulated dataset: 3.1-6.0 dB, 0.92-3.0dB, and -1.6-0.92 dB.  Here the signal noise indicates level of contamination in EEG itself and not labels. In addition to the results showed in main text using the data at 1.6-0.92 dB SNR, here we compare across three SNR levels in Fig (\nameref{S1_snr}). We present results from symmetric label noise condition (rand) by randomly changing seizure annotations between two classes (seizure/ no seizure) in contiguous time windows. This noise configuration was chosen due to its minimal assumptions and its representation of general symmetric noise scenarios.
Our results shown in Fig (\nameref{S1_snr}) demonstrate that the median AUROC for all three models remains consistently high across varying SNR levels when trained with BUNDL. In contrast, models trained using the two baseline methods and CEL exhibit a gradual decline in AUROC as the SNR increases, a trend that is  pronounced in the DeepSOZ and TGCN models.
BUNDL, however, has better performance in all scenarios. Further, the we also observe that FPR significantly increases for baseline methods with decreasing SNR,  with slight decrease in sensitivity. BUNDL on the other hand, achieves the least FPR while maintaining comparable sensitivity.  These findings underscore the robustness of BUNDL in maintaining superior performance despite increased noise interference in the data.}

\paragraph*{S1 Fig 1.}
\label{S1_snr}
{\textbf{Performance metrics at different EEG signal-to-noise ratio levels}} 
{Box plots of AUROC, FPR (min/hour), and sensitivity metrics are shown from three deep networks  trained using BUNDL and baseline strategies at three different EEG signal-to-noise levels and symmetric (rand) noisy training labels. Here the signal noise indicates level of contamination in EEG itself and not labels. }
\begin{figure*}[t]
    \centering
    \includegraphics[width=0.95\columnwidth]{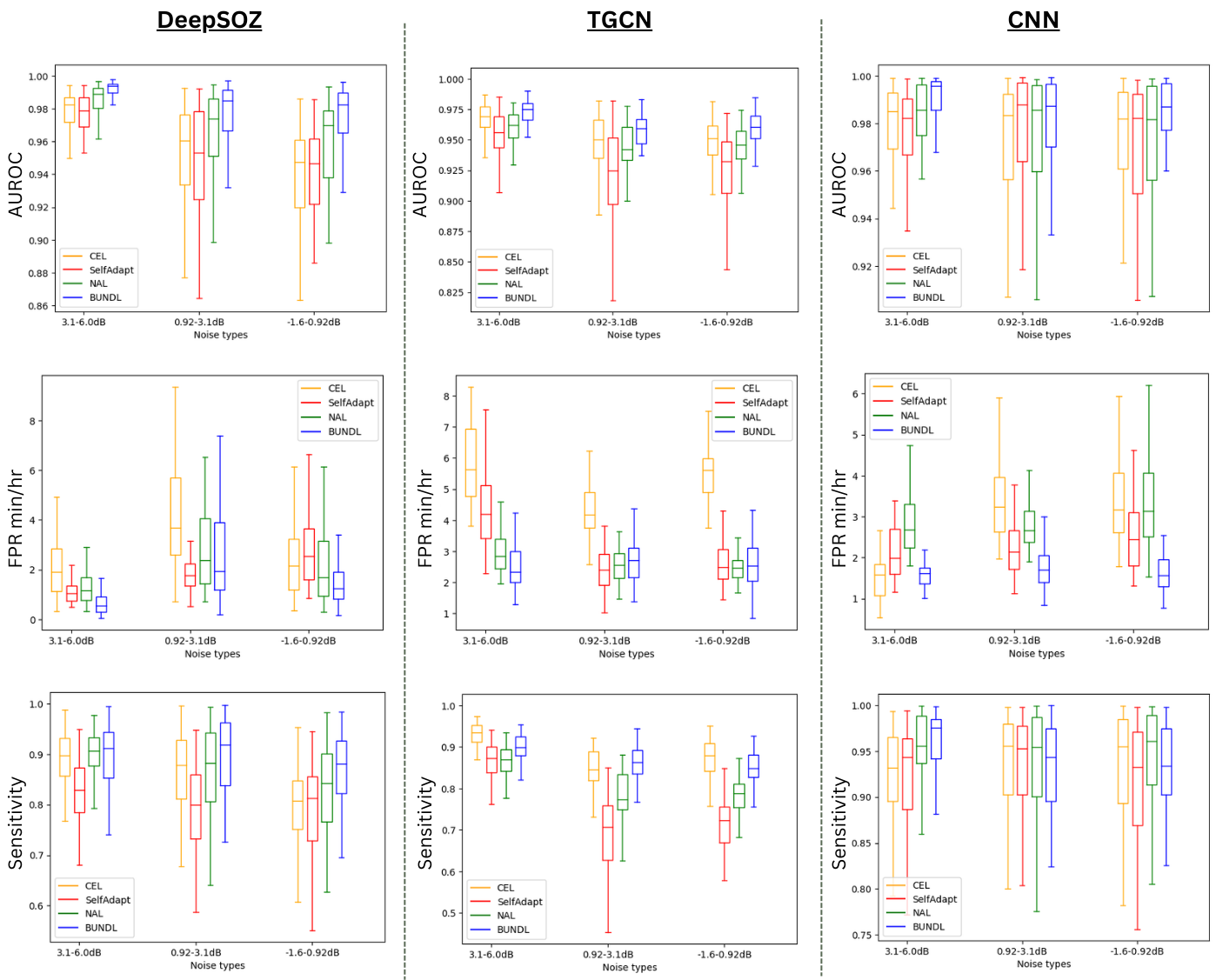}
    \caption{Box plots of AUROC, FPR (min/hour), and sensitivity metrics of three models trained using BUNDL and baseline strategies at three different SNR levels and randomly noisy training labels. }
    \label{S1_snr}
\end{figure*}

%\section*{Acknowledgments}
% This work was supported by the National Institutes of Health R01 EB029977 (PI Caffo),  R01 HD108790 (PI Venkataraman), R21 CA263804 (PI Venkataraman), and the National Science Foundation CAREER 1845430 (PI Venkataraman).

\nolinenumbers

\bibliography{main} 

\end{document}